\documentclass[twocolumn]{emulateapj}

\usepackage{times}
\usepackage{amsmath,amssymb,amstext}
\usepackage[colorlinks=blue,linkcolor=blue,urlcolor=blue,citecolor=blue]{hyperref} 
\usepackage[all]{hypcap}
\usepackage{tabularx}
\usepackage{float}
\usepackage{natbib}
\usepackage{longtable}
\usepackage[usenames, dvipsnames]{color}
\usepackage[switch]{lineno}

\begin{document}

\title{Exploring the connection between parsec-scale jet activity and broadband outbursts in 3C 279}
\author{B. Rani\altaffilmark{1}, S. G. Jorstad\altaffilmark{2,7}, A. P. Marscher\altaffilmark{2},   I. Agudo\altaffilmark{3}, K. V. Sokolovsky\altaffilmark{4,5,6}, V.~M.~Larionov\altaffilmark{7,8}, P. Smith\altaffilmark{9}, D. A. Mosunova\altaffilmark{5}, G.~A.~Borman\altaffilmark{10}, T.~S.~Grishina\altaffilmark{7},  
E.~N.~Kopatskaya\altaffilmark{7}, 
A.~A.~Mokrushina\altaffilmark{7,8}, 
D.~A.~Morozova\altaffilmark{7}, 
S.~S.~Savchenko\altaffilmark{7}, 
Yu.~V.~Troitskaya\altaffilmark{7}, 
I.~S.~Troitsky\altaffilmark{7}, 
C. Thum\altaffilmark{11}, 
S. N. Molina\altaffilmark{3},
C. Casadio\altaffilmark{12, 3}  
}
\affil{$^1$NASA Goddard Space Flight Center, Greenbelt, MD, 20771, USA}
\affil{$^2$Institute for Astrophysical Research, Boston University, 725 Commonwealth Avenue, Boston, MA 02215, USA }
\affil{$^3$Instituto de Astrof\'{\i}sica de Andaluc\'{\i}a (CSIC),  
                 Apartado 3004, E--18080 Granada, Spain}   
\affil{$^4$IAASARS, National Observatory of Athens, Vas.~Pavlou \& I.~Metaxa, GR-15~236 Penteli, Greece}
\affil{$^5$Astro Space Center of Lebedev Physical Institute, Profsoyuznaya~St.~84/32, 117997~Moscow, Russia}
\affil{$^6$Sternberg Astronomical Institute, Moscow State University, Universitetskii~pr.~13, 119992~Moscow, Russia}
\affil{$^7$Astron.\ Inst., St.-Petersburg State Univ., Russia}
\affil{$^8$Pulkovo Observatory, St.-Petersburg, Russia}
\affil{$^9$Steward Observatory, University of Arizona, Tucson, AZ 85721, USA}
\affil{$^{10}$Crimean Astrophysical Observatory, P/O Nauchny, Crimea, 298409, Russia} 
\affil{$^{11}$Instituto de Radio Astronom\'ia Millim\'etrica, 
                Avenida Divina Pastora, 7, Local 20, E--18012 Granada, Spain }
\affil{$^{12}$Max--Planck--Institut f\"ur Radioastronomie, 
                Auf dem H\"ugel, 69, D--53121, Bonn, Germany }
\thanks{NPP Fellow}
\email{bindu.rani@nasa.gov}
\begin{abstract}
We use a combination of high-resolution  very long baseline interferometry (VLBI) radio 
and multi-wavelength flux density and polarization observations  to 
constrain the physics of the dissipation mechanism powering the broadband flares 
in 3C~279 during an episode of extreme flaring activity in 2013-2014. Six bright flares 
superimposed on a long-term outburst are detected at $\gamma$-ray energies. Four of the flares have optical 
and radio counterparts. The two modes of flaring 
activity (faster flares sitting on top of a long term outburst) present at 
radio, optical, $\gamma$-ray frequencies are missing in X-rays. X-ray counterparts 
are only observed for two flares. The first three flares are accompanied by ejection 
of a new VLBI component (NC2) suggesting the 43 GHz VLBI core as the site of energy dissipation. 
Another new component, NC3, is ejected after the last three flares, which suggests that 
the emission is produced  upstream from the core (closer to the black hole). The 
study therefore indicates multiple sites  of energy dissipation in the source.   
An anti-correlation is detected  between the optical percentage polarization (PP) and 
optical/$\gamma$-ray flux variations, while the PP has a positive correlation with 
optical/$\gamma$-rays spectral indices. Given that the mean polarization is  
inversely proportional to the number of cells in  the emission region, the PP 
vs.\ optical/$\gamma$-ray anti-correlation could be due to more active cells 
during the outburst than at other times. In addition to the turbulent component, 
our analysis suggests the presence of a combined turbulent and ordered magnetic field, 
with the ordered component transverse to the jet axis.
\end{abstract}


\keywords{galaxies: active -- quasars: individual: 3C~279 -- 
             radio continuum: galaxies -- jets: galaxies -- gamma-rays  
               }


\section{Introduction}
Combined information from the high-resolution images obtained using  
very long baseline interferometry (VLBI) radio and broadband intensity variations has been 
proven  
to be one of the most useful approaches to locate and to probe the high-energy radiation 
processes in 
blazars -- Active Galactic Nuclei (AGN) in which one of the jets points close to our line of sight. 
Investigation of intensity variations in individual sources in combination with the VLBI 
component ejection/kinematics indicates that high-energy emission is associated with the 
compact regions in relativistic jets \citep[e.g.][and references therein]{agudo2011, 
jorstad2010, marscher2008, rani2014, rani2015, rani2016}. 
A recent study for blazar S5 0716+714 by \citet{rani2014} reported a significant correlation 
between the $\gamma$-ray flux and the direction of the jet outflow variations. This study implies 
a causal connection between the jet morphology and the observed $\gamma$-ray flares.

We present here an investigation of parsec-scale jet morphology evolution in the blazar 
3C~279 during an episode of extreme flaring activity at GeV energies in 2014, using 43~GHz 
VLBI images. 
The focus of this study 
is to  explore the total intensity and polarization properties and their connection with the 
broadband flux variations observed during this bright $\gamma$-ray activity phase of the source.

The flat spectrum radio quasar (FSRQ) 3C 279, at a redshift of z = 0.538 \citep{burbidge1965}, 
has a black hole mass in the range of (3--8)$\times$10$^8$~M$_{\odot}$ \citep{woo2002, gu2001}.
3C~279 has a bright jet extending up to kiloparsec scales \citep{cheung2002} and exhibits a range of apparent 
velocities from 4 to 20~$c$ \citep{lister2013}. The parsec scale jet is aligned close to the observer's 
line of sight \citep[$\leq$2$^{\circ}$,][]{lister2013, jorstad2004}. Because of its strong optical polarization and flux variability, 3C~279 is often classified as an 
optically violently variable (OVV) quasar. Polarimetric observations have detected both linearly and circularly 
polarized emission from the parsec-scale jet of 3C 279 \citep{wardle1998, taylor2000, 
homan2009}. Moreover, the observed radiation at optical frequencies is also 
highly polarized in the source e.g.\ an observed value of 45.5$\%$ in optical 
U band \citep{mead1990}. \citet{wagner2001} detected variable optical circular polarization in 3C~279 
exceeding 1$\%$.

With its  high flux density and prominent variations in total  intensity and 
polarization, 3C~279 has been the subject of several intensive multi-wavelength campaigns 
\citep[e.g.,][]{maraschi1994, hartman1996,  chatterjee2008, wehrle1998, larionov2008, collmar2010, 
bottcher2007, hayashida2012, hayashida2015, kang2015}.  In particular, prominent $\gamma$-ray 
flares were recorded in the source by the 
Energetic Gamma-Ray Experiment Telescope (EGRET) on board the {\it Compton Gamma-Ray 
Observatory (CGRO)} \citep{hartman1992}, and  the {\it Fermi}/LAT (Large Area Telescope) has detected the 
source  regularly since its launch in 2008 \citep{abdo2010a}. In 2006, the Major Atmospheric Gamma-Ray 
Imaging Cherenkov (MAGIC) telescope observed TeV (E$>$200~GeV) emission from the source \citep{albert2008}. 
In December 2013, the source went through a series of extremely rapid $\gamma$-ray flares reaching 
a flux level F (E $>$100~MeV) of 10$^{-5}$ photons cm$^{-2}$ s$^{-1}$ on timescales of the order of 
a few hours \citep{hayashida2015}. In June 2015,  record-breaking flaring activity was observed \citep{cutini2015} 
with the highest measured flux, F (E $>$100~MeV) of 3.6$\times$10$^{-5}$ 
photons cm$^{-2}$ s$^{-1}$ \citep{fermi_3c279_2016}.

 This paper is structured as follows. Section 2 provides a brief description of observations and data 
reduction. In Section 3, we report and discuss our results. Finally, summary and conclusions are given 
in Section 4. 
Throughout this paper we adopted a $\Lambda$CDM cosmology with 
$\Omega_m$ = 0.27, $\Omega_{\lambda}$ = 0.73, 
and $H_0$ = 71 km s$^{-1}$ Mpc$^{-1}$ \citep{spergel2003}.  
At a redshift of 0.538, the luminosity distance $d_L$ 
is  3085~Mpc and an angular separation of 1 milli-arcsecond (mas) 
translates into a linear distance of 6.32~pc.



\section{Multifrequency data: observations and data reduction}
From  November 01, 2013 to August 30, 2014, the broadband flaring activity of the FSRQ 
3C 279 was extensively covered using both ground- and space- based observing facilities. 
The following sub-sections summarize the observations and data reduction.

\subsection{VLBA Observations}
The 7~mm VLBA observations of the source were obtained in the course of
a program\footnote{VLBA-BU-BLAZARS, http://www.bu.edu/blazars} of monthly 
monitoring of bright $\gamma$-ray blazars at 43 GHz. The details of the 
observations can be found in \citet{jorstad2017}. The data reduction 
is performed using the Astronomical Image Processing System (AIPS) and
Difmap (Shepherd 1997). Details of the data reduction are described in 
\citet{jorstad2005, jorstad2017}.  We model-fit the brightness distribution of the source 
using circular Gaussian components using Difmap, and the model-fit parameters 
(peak flux density, distance from (0,0) co-ordinates, position angle, and 
size) are listed in the  Table \ref{tab1} in Appendix. The polarization 
parameters (polarized flux density and electric vector position angle, EVPA) 
are determined by model-fitting the stokes Q and U in 
({\it u,v}) data. The polarized flux density is given by 
$\sqrt{Q_\mathrm{flux}^2 + U_\mathrm{flux}^2}$ and the  EVPA by 
0.5$\times$$\arctan(U_\mathrm{flux}/Q_\mathrm{flux})$.

\subsection{Radio observations} 

The 230 and 345 GHz data are provided by the Submillimeter Array (SMA) Observer
Center\footnote{http://sma1.sma.hawaii.edu/callist/callist.html} data base \citep{gurwell2007}. 
The 230 GHz total  flux density data  are complemented by the data from the POLAMI 
(Polarimetric Monitoring of Active Galactic Nuclei at Millimetre Wavelengths) 
Program\footnote{\href{url}{https://polami.iaa.es}} \citep{agudo2017a, agudo2017b, thum2017}. 
POLAMI is a long-term program to monitor the polarimetric properties 
(Stokes I, Q, U, and V) of a sample of around 40 bright AGN 
at 3.5 and 1.3 millimeter wavelengths with the IRAM 30m Telescope near Granada, Spain. 
The program has been kept running since October 2006, and it currently has a time sampling  
of $\sim$2 weeks. The XPOL polarimetric observing setup has been 
routinely used as described in \citet{thum2008} since the start of the program. 
The reduction and calibration of the POLAMI data presented here are described in 
detail in \citet{agudo2010, agudo2014, agudo2017a}.  
We also make use of the 15~GHz OVRO (Owens Valley Radio Observatory) data 
publicly available at the OVRO website\footnote{http://www.astro.caltech.edu/ovroblazars/}. Details of 
observations and data reduction are described in  \citet{richards2011}.

\subsection{Optical observations}

During our campaign period, the source was simultaneously observed in 6 optical/UV filters 
(U, B, V, W1, W2, and M2) by {\it Swift}/UVOT \citep{roming2005}. 
We used the {\it uvotsource} tool from HEASoft~v6.16 for the data analysis
\citep{poole2008}. A step-by-step data reduction procedure is described in \citet{rani2017_3c279}  
to which we refer for details. 
 We have also included optical/IR data at B, V, R, J, and K passbands from the Small and Moderate Aperture Research Telescope 
System (SMARTS5) monitoring program. The reduction and analysis of this data is described in 
\citet{bonning2012}.

We also make use of the optical R passband flux and polarization 
observations  provided by the Steward observatory blazar monitoring 
program\footnote{http://james.as.arizona.edu/~psmith/Fermi/}; the details of 
observations and data reduction are described in \citet{smith2016}. 
Additional optical R band polarimetric observations were obtained at the 1.83~m Perkins 
telescope of the Lowell Observatory (Flagstaff, AZ) using
the PRISM camera\footnote{www.bu.edu/prism/}. The 
details of observations and data reduction are given in  \citet{jorstad2016}. 
We also used polarimetric and photometric data from the AZT-8 (Crimean Astrophysical Observatory) 
and LX-200 (St.Petersburg, Russia)    telescopes. The details of 
observations and data reductions are given in \citet{larionov2008}.

Finally, data were corrected for the Galactic extinction using the
coefficients of \citet{schlegel1998} for different filters\footnote{search for 3C 279 at http://ned.ipac.caltech.edu/}.
We converted the magnitudes to fluxes using the central wavelengths for each filter as
calibrated by \citet{poole2008} for the space-based observations. For ground-based observations, 
the zero-point fluxes\footnote{http://ssc.spitzer.caltech.edu/warmmission/propkit/pet/magtojy/ref.html} 
corresponding to the observed wavelengths were used for magnitude to flux conversion.

\subsection{X-ray observations}

X-ray observations (0.3 $-$ 10\,keV) of the source were obtained using the Swift/XRT 
telescope \citep{burrows2005} both in the photon counting and windowed timing mode. 
The data are analyzed 
using the standard xrtpipeline v0.13.2. The spectra in individual 
time bins were modeled with XSPEC~v12.9.0 to extract the photon counts and indices. 
We used the absorbed power law with the total HI column 
density fixed to the Galactic value \citep[$N_{HI} = 0.212\times10^{21}$,][]{kalberla2005} 
to model the 0.3 $-$ 10\,keV spectrum, which provided an acceptable
fit  at all epochs. The details of the observations and data 
reduction are discussed in \citet{rani2017_3c279}.

  \begin{figure*}
\includegraphics[scale=1.2,angle=0, trim=0 0 0 0.5, clip]{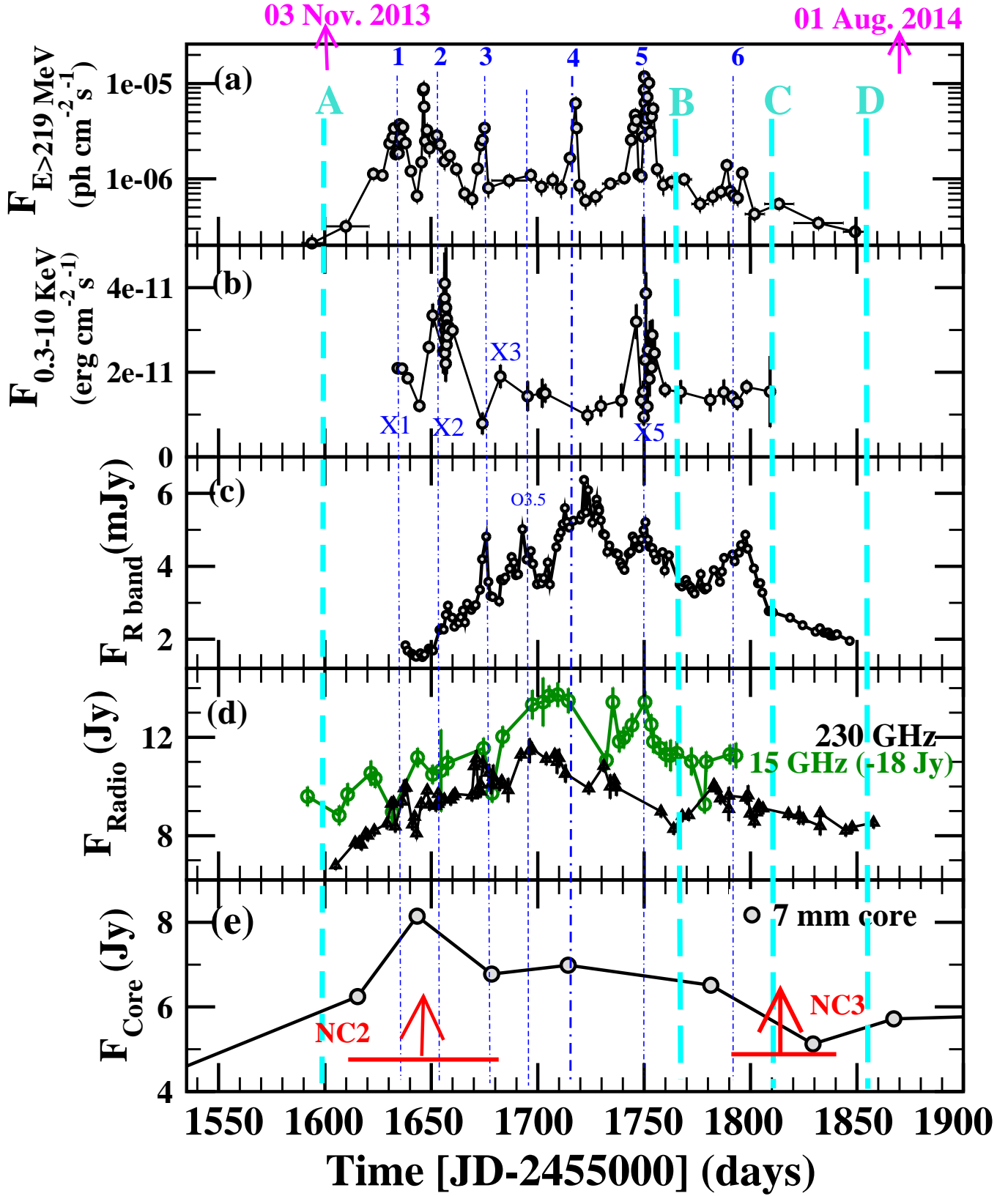}
   \caption{  Broadband flux light curves of 3C 279: (a) $\gamma$-ray light curve at E$>$219~MeV; (b) 
X-ray light curve at 0.3-10~keV; (c) optical R passband light curve, (d) 230 and 15~GHz radio flux density light 
curves, and (e) flux density variations in the 7~mm VLBI core. 
The red arrows mark the ejection times of new components (see Sect.\ \ref{sect_vlbi} for details). As for other figures hereafter,  the cyan dashed lines (labelled as A to D) represent the different 
flaring activity states of the source (see Sect.\ \ref{sect_lc}  for 
more details). The peaks of the rapid $\gamma$-ray flares (1 to 6) are marked with the blue dotted-dashed lines. The orange dashed line shows the time of an optical flare (3.5) that has no obvious counterparts at other wavelengths. In all cases, lines connecting the data points are simply to guide the eye. }
\label{plot_lcs}
\end{figure*}

\subsection{Gamma rays}
We employed here the 100~MeV -- 300~GeV data of the source observed in survey mode by the {\it Fermi}-LAT 
\citep[Large Area Telescope,][]{atwood2009}. The LAT data were analyzed using the standard ScienceTools (software 
version v10.r0.p5) and instrument response function P8R2$\_$SOURCE$\_$V6. Photons in the Source event class were 
selected for the analysis. We analyzed a region of  interest (ROI) of 20$^{\circ}$, centered at the position
of 3C~279 using a maximum-likelihood algorithm \citep{mattox1996}.  For the unbinned likelihood 
analysis\footnote{http://fermi.gsfc.nasa.gov/ssc/data/analysis/scitools/likelihood\_tutorial.html}, 
all 3FGL \citep{3fgl_paper} sources within the ROI were included. We also included the Galactic diffuse 
background ({\it $gll\_iem\_v06.fits$}) and the isotropic background ({\it $iso\_P8R2\_SOURCE\_V6\_v06.txt$}) emission 
components\footnote{https://fermi.gsfc.nasa.gov/ssc/data/access/lat/BackgroundModels.html}. 
Model parameters for sources within 5$^{\circ}$  of the center of the  ROI were 
kept free. Model parameters for variable sources (variability index $\geq$72.44 in the 3FGL catalog) were also 
set free while the model parameters
of non-variable sources beyond 5$^{\circ}$ were fixed to their catalog values. 

To characterize the variability properties of the source, we computed the photon flux light curve of the 
source at E above E$_0$, where E$_0$ is the  de-correlation energy \citep{lott2012}, which minimizes the 
correlations between integrated photon flux and photon index. During  our campaign, we obtain 
E$_0$ = 219~MeV. We generated the constant uncertainty (15$\%$) light curve above E$_0$ through the 
adaptive binning method following \citet{lott2012}.  The bin size ranges between 0.07 to 31.06~days with and average of 5.43~days. 
The adaptive binned light curve is produced by 
modeling the spectra by a simple power law, N(E) = N$_0$ E$^{-\Gamma}$, N$_0$ : prefactor, 
and $\Gamma$ : power-law index.


\section{Results and discussion} 
This section presents the light curve analysis -- flux and polarization variability at 
different energy bands and the evolution of the jet brightness. We also investigate 
the connection between jet kinematics and broadband flaring activity.

\subsection{Observed outburst and its multi-frequency behavior}
\label{sect_lc}

The observed broadband flux density variations in 3C~279 during a major $\gamma$-ray outburst are plotted in 
Fig.\ \ref{plot_lcs}.   The source became active 
at GeV energies in early November 2013, and the flaring activity 
continued  until August 2014. Hereafter we will use JD$^{\prime}$ (defined as JD - 2455000) for time.
In Fig.\ \ref{plot_lcs}, the cyan dashed lines labeled as ``A" and ``D" mark the 
beginning and end of the outburst, respectively. 
The source displayed multiple flares across the whole electromagnetic spectrum over this period, 
which we visually inspected and labeled 
as follows.  The blue dotted-dashed lines (``1" to ``6") are  marked close to the peak of 
the rapid $\gamma$-ray flares labeling the 
broadband flares as ``G" for $\gamma$-rays, ``X" for X-rays, ``O" for optical, ``R" for radio  followed by the 
number. For instance, the $\gamma$-ray flares should be read as G1 to G6.

\begin{figure}
\includegraphics[scale=0.35,angle=0, trim=0 0 0 0.0, clip]{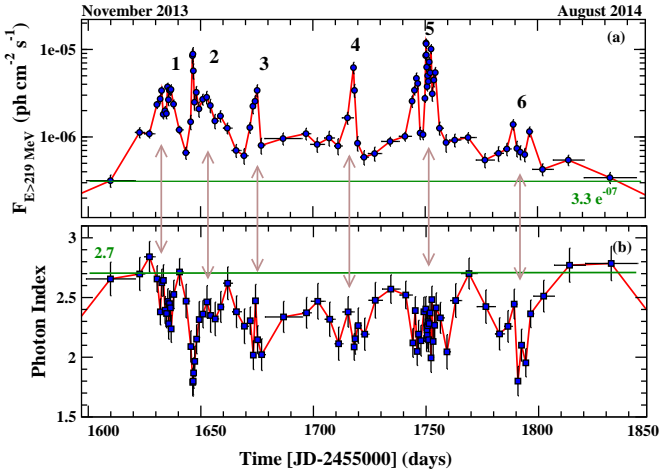}
   \caption{Gamma-ray photon flux (top) and photon index (bottom) light curves of 3C~279. The bright $\gamma$-ray 
flare and the corresponding spectral variations are shown with the gray arrows. The photon flux and photon index 
values  at the beginning and end of the outburst are marked with the green lines.  }
\label{plot_fig1}
\end{figure}

Flare 2 is 
observed close in time at $\gamma$-ray and X-ray energies.  The first 
two flares are accompanied by  an increase in the source brightness at radio frequencies. At optical frequencies (panel c) 
the source is still in quiescent phase.  
The peak of flare G3 coincides with a minimum in the X-ray light curve, but has  optical and radio 
counterparts and a delayed response in X-rays. In comparison to X1 and X2, X3 has a  rather low magnitude.  
Because of the gap in observations at X-ray energies, it is hard to determine the absence/presence of an X-ray counterpart for 
flare G4, but the flare has a near-simultaneous optical and radio counterpart. 
A prominent optical flare (labeled as 3.5) is seen between 
flare 3 and 4. The flare O3.5 is accompanied by  similar activity at radio (230 GHz) band; however, the observed variations are 
quite mild at X-ray and $\gamma$-ray energies.  
Flare 5 is the only one that is observed quasi-simultaneously 
at all energies (radio to $\gamma$-rays).  Flare 5 has a double 
feature in $\gamma$-rays and X-rays (see Figs.\ \ref{plot_fig1} and \ref{plot_fig2}).

The other two interesting states in the broadband activity are marked as ``B" and ``C".  After the first five $\gamma$-ray 
flares, a local minimum (``B")  is observed simultaneously at all frequencies. The local minimum was followed by a concurrent 
flare (flare 6) at radio, optical, $\gamma$-ray energies; no significant activity was observed in the X-ray regime during 
this period. Period ``C" marks the end of flare 6, which happened to be almost simultaneous for all  light curves. A 
continuous decay in the source brightness was observed between period C and D, which we refer to as the post-outburst 
plateau.

  \begin{figure}
   \centering
\includegraphics[scale=0.35,angle=0, trim=0 0 0 0.0, clip]{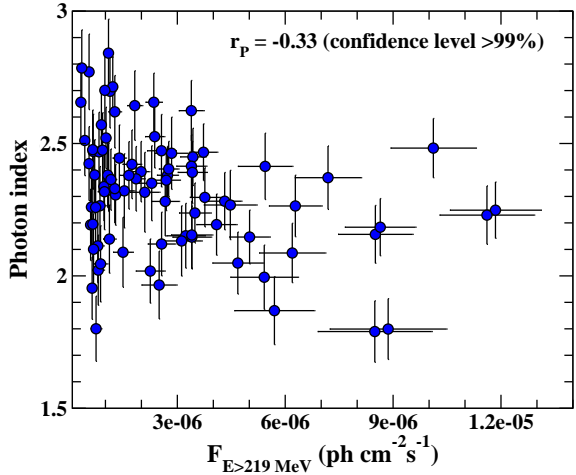}
   \caption{Gamma-ray photon index vs.\ photon flux for E$>$219~MeV. }
\label{plot_fig_gamma}
\end{figure}

\subsubsection{Gamma-ray Variability}
\label{lc_gamma}
Figure \ref{plot_fig1} displays details of the observed $\gamma$-ray flaring activity of 3C~279.
The rapid flares (labeled as 1 to 6) were superimposed on top of a broad outburst. 
 The source started to brighten from a flux level 
F (E $>$219~MeV) $\sim$ 3.3$\times$10$^{-7}$ photons cm$^{-2}$ s$^{-1}$ on JD$^{\prime}$ $\sim$1610  and 
 returned to the same brightness level on JD$^{\prime}$ $\sim$1850. The solid green line in Fig.\ \ref{plot_fig1} marks 
the observed brightness level ($\sim$ 3.3$\times$10$^{-7}$ photons cm$^{-2}$ s$^{-1}$)  of the source 
before and after the outburst period.  
Significant variations in $\gamma$-ray photon index were observed during this period.  The estimated $\Gamma$ value is 2.7  at the beginning and end
of the outburst. A significant spectral variation is observed for the individual flares (1 to 6). For each flare, 
the spectrum gets harder at the peak, and later it softens again. The fastest observed $\gamma$-ray flare during 
this activity period had a flux-doubling time scale of $\sim$2~hrs and a very hard $\gamma$-ray spectrum with 
$\Gamma$ = 1.7 \citep{hayashida2015}. 

In Fig.\ \ref{plot_fig_gamma}, we plot the $\gamma$-ray photon flux values as a 
function of photon index. A clear spectral hardening 
with an increase in source brightness can be seen here. 
 Since we investigate the correlation above the de-correlation 
energy (E$_{0}$), the estimated correlation is not sensitive to the LAT instrumental bias.
A Pearson correlation analysis reveals a significant
correlation between the two. Formally, we obtained $r_P$\footnote{$r_P$ is the linear Pearson correlation coefficient 
and p-value is the probability of the null hypothesis} = $-$0.33 and a p-value = 0.0019, which suggests a 
significant anti-correlation at a confidence level of $>$99$\%$. It is therefore evident that not only the major 
outburst but also the individual $\gamma$-ray flares are characterized by significant spectral hardening although 
with substantial scatter.

  \begin{figure}
\includegraphics[scale=0.35,angle=0, trim=0 0 0 0.5, clip]{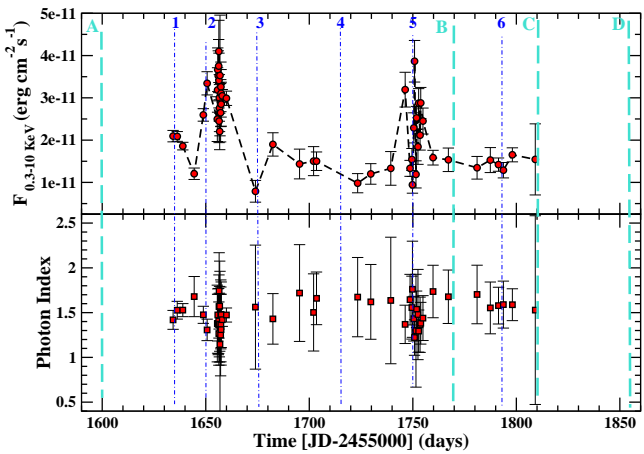}
   \caption{Energy flux (top) and photon index (bottom) light curves of 3C~279 at 0.3--10~KeV X-ray 
bands (vertical dotted and dashed lines are same as in Fig.\ 1). }
\label{plot_fig2}
\end{figure}

\subsubsection{X-ray Variability}
The X-ray energy flux and photon index light curves are shown in Fig.\ \ref{plot_fig2}. 
Compared to the $\gamma$-ray light curve,  the X-ray observations were rather sparse.  We 
could identify two prominent flares in the X-ray light curve. Apparently, the first X-ray flare  
coincides with  flare G2 in $\gamma$ rays and  the second X-ray flare with flare G5. While the X-ray energy flux 
is varying by a factor of more than four, the photon index remains consistent with the mean 
value of $\Gamma_{X} = 1.47 \pm 0.01$.

  \begin{figure}
\includegraphics[scale=0.5,angle=0, trim=0 0 0 0.5, clip]{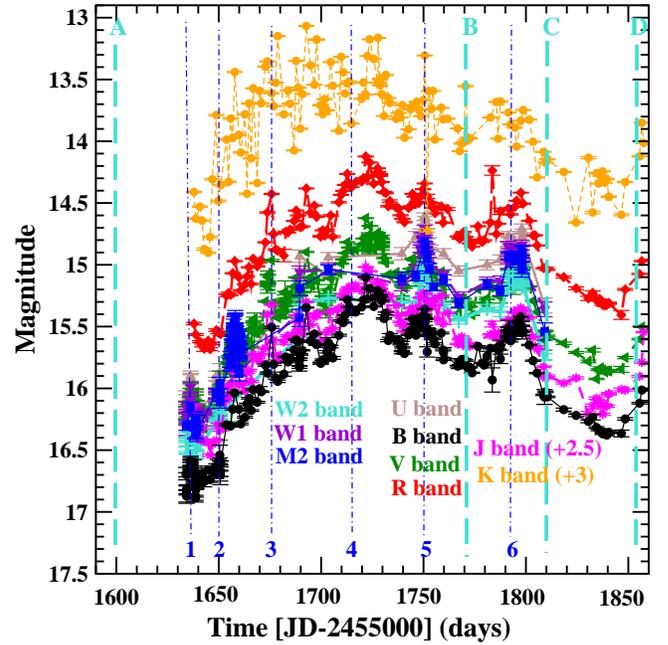}
   \caption{Near-IR (J and K), UV (W1, W2, and M2), and optical (U, B, V, and R) light curves of 
3C~279 (vertical dotted and dashed lines are same as in Fig.\ 1).   }
\label{plot_fig3}
\end{figure}

\subsubsection{Variability at Optical frequencies}
\label{lc_opt}
The observed near-IR, optical, and UV light curves of the source are shown in Fig.\ \ref{plot_fig3}. 
Similar to $\gamma$-rays, there are two modes of flaring activity at optical frequencies. Fast 
flares (timescale $\sim$ 20-30~days) are superimposed on the broad outburst. No optical observations are available 
before JD$^{\prime}$ = 1635 (December 2013).  The flaring behavior is very similar for the near-IR, optical, 
and UV light curves.

  \begin{figure}
\includegraphics[scale=0.35,angle=0, trim=0 0 0 0.5, clip]{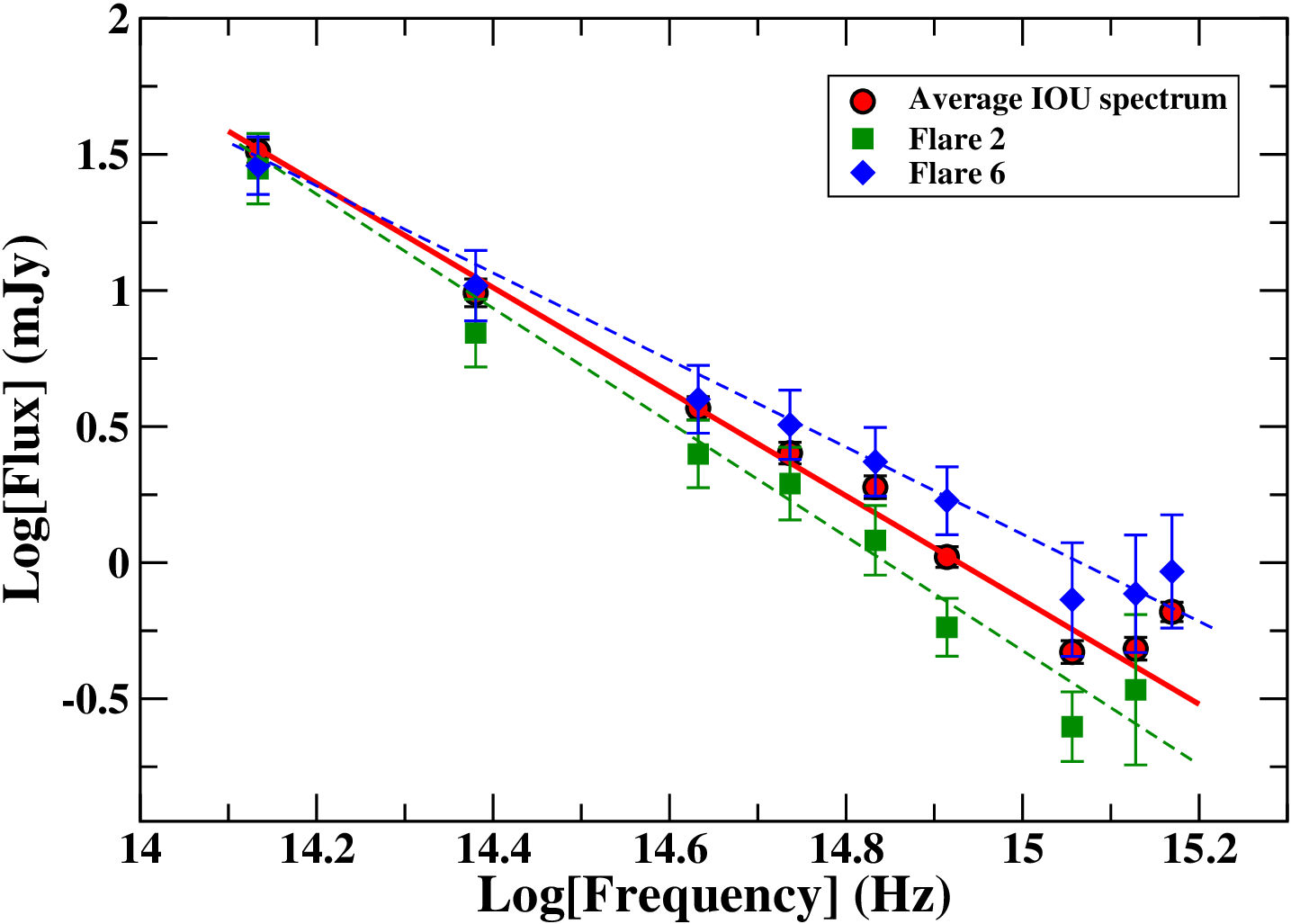}
\includegraphics[scale=0.35,angle=0, trim=0 0 0 0.5, clip]{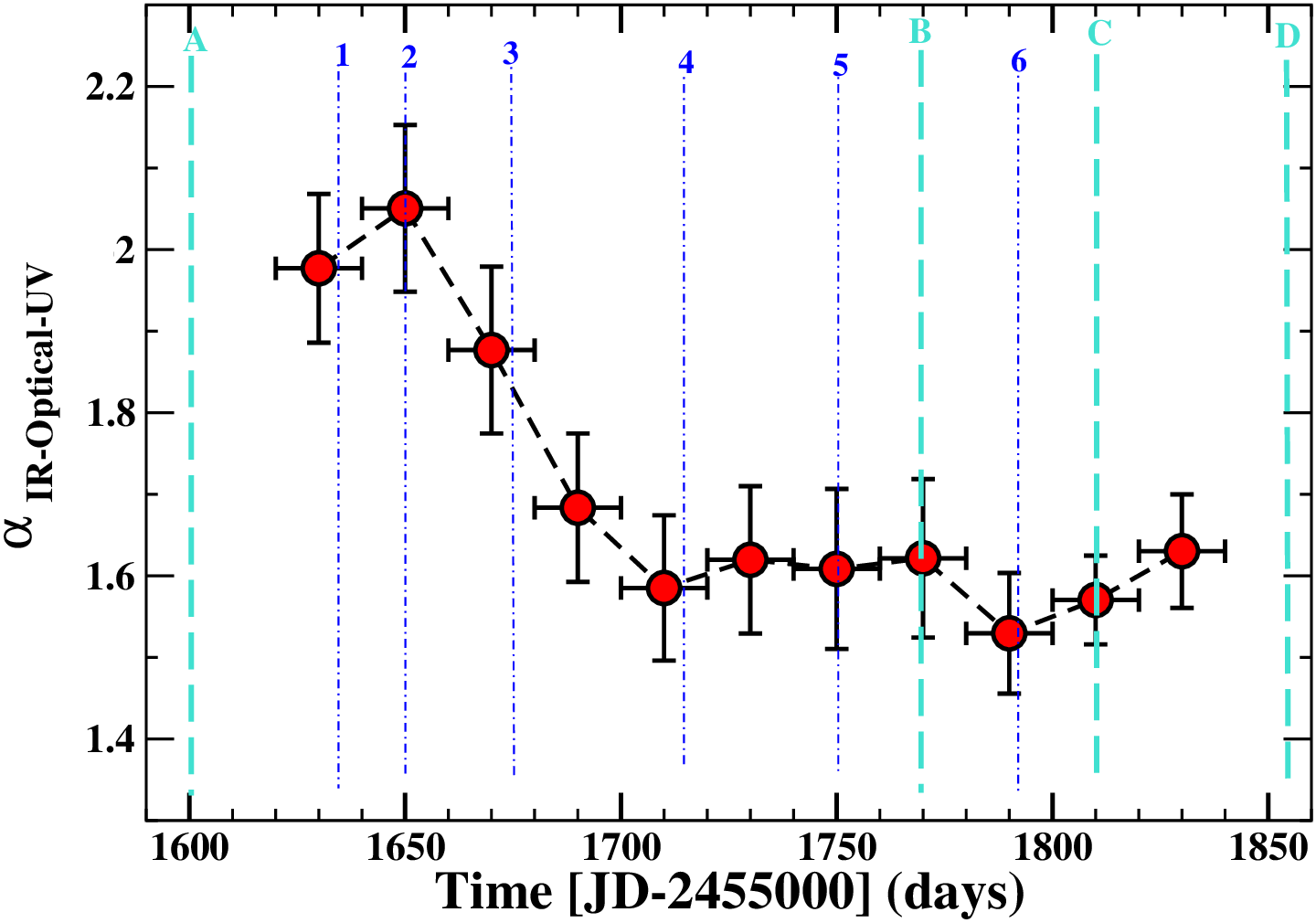}
   \caption{Top: Average IOU spectrum  (data points are the mean values 
and the error bars representation their standard deviation)} of 3C~279 over the entire 
period of observations, and 
a change  in spectral index from $\sim$2 (Flare 2 in green) to $\sim$1.6 (Flare 6 in blue). 
Bottom: Spectral index variations (vertical dotted and dashed lines are same as in Fig.\ 1).
\label{plot_fig3a}
\end{figure}

The average IR-optical-UV (hereafter IOU) spectrum over the entire period of our observations is shown in 
Fig.\ \ref{plot_fig3a} (top). The observed IOU spectrum  is steep  until an upturn at UV frequencies. The excess emission could be due to  thermal dominance either from the accretion disk or broad-line region.   
The observed IOU spectrum can be well described  by  a power law, F($\nu$) $\propto$ $\nu^{-\alpha}$, 
where $\alpha$ is the spectral index.  For the average spectrum, we obtained $\alpha$ = 1.91$\pm$0.08.

 To investigate the spectral variations during the broad outburst,  we constructed the IOU spectrum 
every 20~days. The spectral index is estimated via fitting the  spectrum by a power law as described above. 
Prominent variations in the IOU spectral indices can be seen in Fig.\ \ref{plot_fig3a} (bottom).  
The IOU spectrum is steeper  at the beginning of the outburst ($\alpha_{IOU} \sim$2, Flare 2), and 
the spectrum gets harder as the source brightness increases. The estimated $\alpha_{IOU}$ is $\sim$1.6 (Flare 6) at the 
peak of the outburst (JD$^{\prime}$ $\sim$ 1720). However, the spectral variations are quite marginal during the decay 
phase of the outburst, with $\alpha_{IOU}$ remaining nearly consistent with a value 
of $\sim$1.6.

Figure 7 presents the optical polarization variations observed in the source. The fractional 
polarization drops from $\sim$25$\%$ to $\sim$5$\%$ within the first 50 days of our observations; 
this period coincides with the beginning of the $\gamma$-ray outburst. The sharp drop in 
percentage polarization was followed by a slow rise (superimposed with rapid variations) until 
August 2014. At the end of the outburst in August 2014, the percentage polarization returned 
to its original value of $\sim$25$\%$. The optical polarization angle (EVPA, $\chi$) variations on the other 
hand are quite modest 
($\Delta \chi$ $\sim$60$^{\circ}$). 
Throughout the campaign
period, EVPA remained consistent with an average value of $\sim$50$^{\circ}$, which is along the parsec-scale jet direction of $\sim$ $-$135$^{\circ}$ \citep{jorstad2017}.

  \begin{figure}
\includegraphics[scale=0.35,angle=0, trim=0 0 0 0.5, clip]{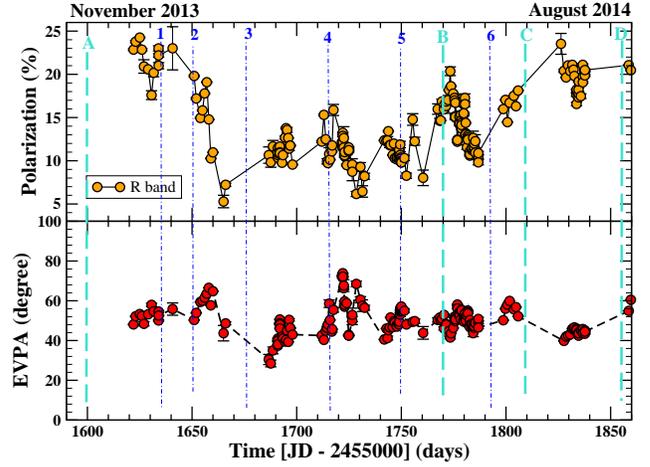}
\caption{Optical R band percentage polarization and EVPA  variations 
(vertical dotted and dashed lines are same as in Fig.\ 1). }
\label{plot_fig3b}
\end{figure}

\subsubsection{Variability at Radio frequencies}
\begin{figure}
\includegraphics[scale=0.41,angle=0, trim=0 0 0 0.5, clip]{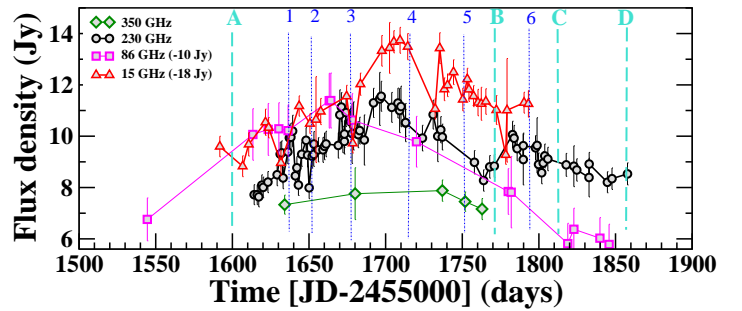}
  \caption{Flux density light curves at 15, 86, 230, and 350~GHz radio bands 
(vertical dotted and dashed lines are same as in Fig.\ 1). The 15~GHz and 86~GHz data shifted 
by $-$18~Jy and $-$10~Jy, respectively.  }
\label{plot_fig4a}
\end{figure}

In Fig.\ \ref{plot_fig4a}, we plot the radio flux density light curves sampled at 15, 86, 
230, and 350~GHz bands. The observed variations appear quite similar at 15 and 230~GHz 
bands, the most densely sampled frequencies. 
 Faster variations (for instance the flare between 3 and 4) superimposed on the broad 
outburst can be seen here. A prominent 
flare is also apparent in the 86~GHz light curve; however data sampling is quite sparse close 
to the peak of the outburst. The data sampling is even worse at the 350~GHz, and as a result 
the flaring activity is less evident.

  \begin{figure}
\includegraphics[scale=0.36,angle=0, trim=0 0 0 0.5, clip]{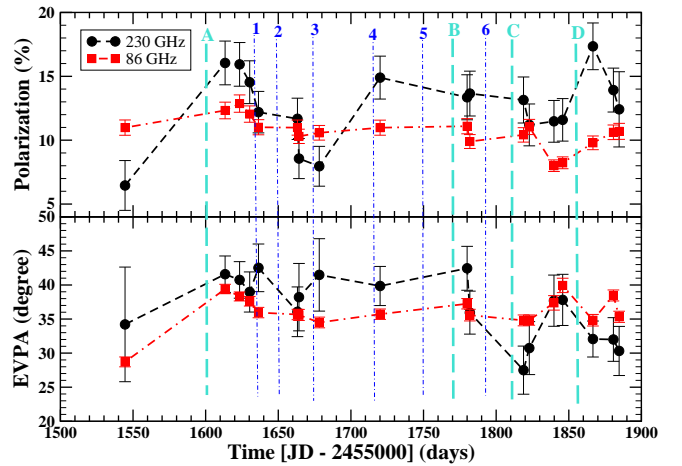}
  \caption{Percentage polarization (top) and EVPA curves at 86 (in red) and 230~GHz 
(in black) bands (vertical dotted and dashed lines are same as in Fig.\ 1). 
 }
\label{plot_fig4b}
\end{figure}

The percentage polarization (top) and EVPA (bottom) curves at 86 and 230~GHz radio bands are shown in 
Fig.\ \ref{plot_fig4b}.  There are significant variations in the 230~GHz band percentage polarization curve. For instance, 
the percentage polarization drops from 15$\%$ to 5$\%$ between JD$^{\prime}$ = 1600 and 1700, and it increases again 
to 15$\%$ at JD$^{\prime}$ $\sim$1720. The observed polarization was quite high ($\sim$15$\%$) for the rest of the 
period. On the other hand, the percentage polarization variations at 86~GHz band and the EVPA variations 
at both 86 and 230~GHz bands were quite modest ($\Delta \chi$ $\leq$25$^{\circ}$).

\subsection{Light curve analysis}

\subsubsection{Fractional Variability}
To quantify how the amplitude of flux variations changes as a function of frequency, we compared the 
fractional variability amplitude ($F_{var}$) of the observed light curves. For a given light curve with 
an average flux $\overline{x}$ and sample variance $S^2$, $F_{var}$ is defined as  
\citep{vaughan2003}: 
\begin{equation}
F_{\rm{var}} = \sqrt{ \frac{S^{2} -
\overline{\sigma_{\rm{err}}^{2}}}{\bar{x}^{2}}}
\end{equation}
where $\overline{\sigma_{\rm{err}}^{2}}$ is the mean of the squared measurement 
uncertainties. The uncertainty on $F_{var}$ due to the measurement-error fluctuations was derived through 
Monte Carlo simulations by \citet[][eq.\ B2]{vaughan2003}.

\begin{figure}
\includegraphics[scale=0.41,angle=0, trim=0 0 0 0.5, clip]{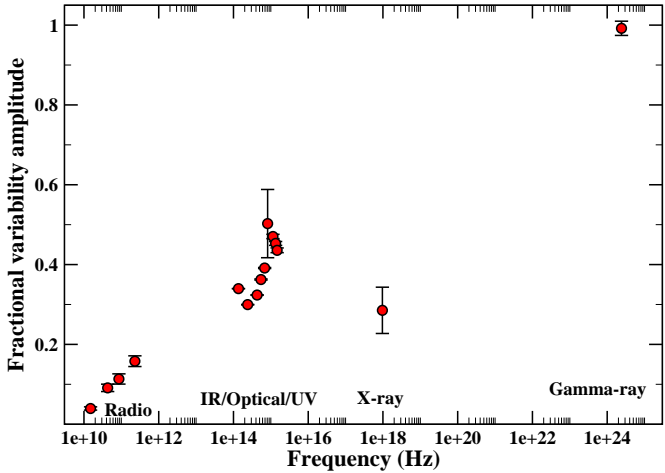}
  \caption{ Fractional variability amplitudes of 3C 279 from radio to $\gamma$-ray energies. }
\label{plot_fig5}
\end{figure}

Figure \ref{plot_fig5} shows the calculated $F_{var}$ at different frequencies. 
A gradual increase in $F_{var}$ with frequency is visible in the 
radio regime, and this trend continues until the IOU regime. The first peak in the 
$F_{var}$ vs.\ frequency curve is observed at the optical U passband, which is followed 
by a dip in the X-ray regime. Compared to the lower frequencies, the $F_{var}$ 
is extremely  large at $\gamma$-rays; $F_{var}$ $>$1 implies more than 100$\%$ variations.

\subsection{Parsec scale jet Kinematics} 
\label{sect_vlbi}
\begin{figure}
   \centering
\includegraphics[scale=0.38,angle=0, trim=0 45 2 80, clip]{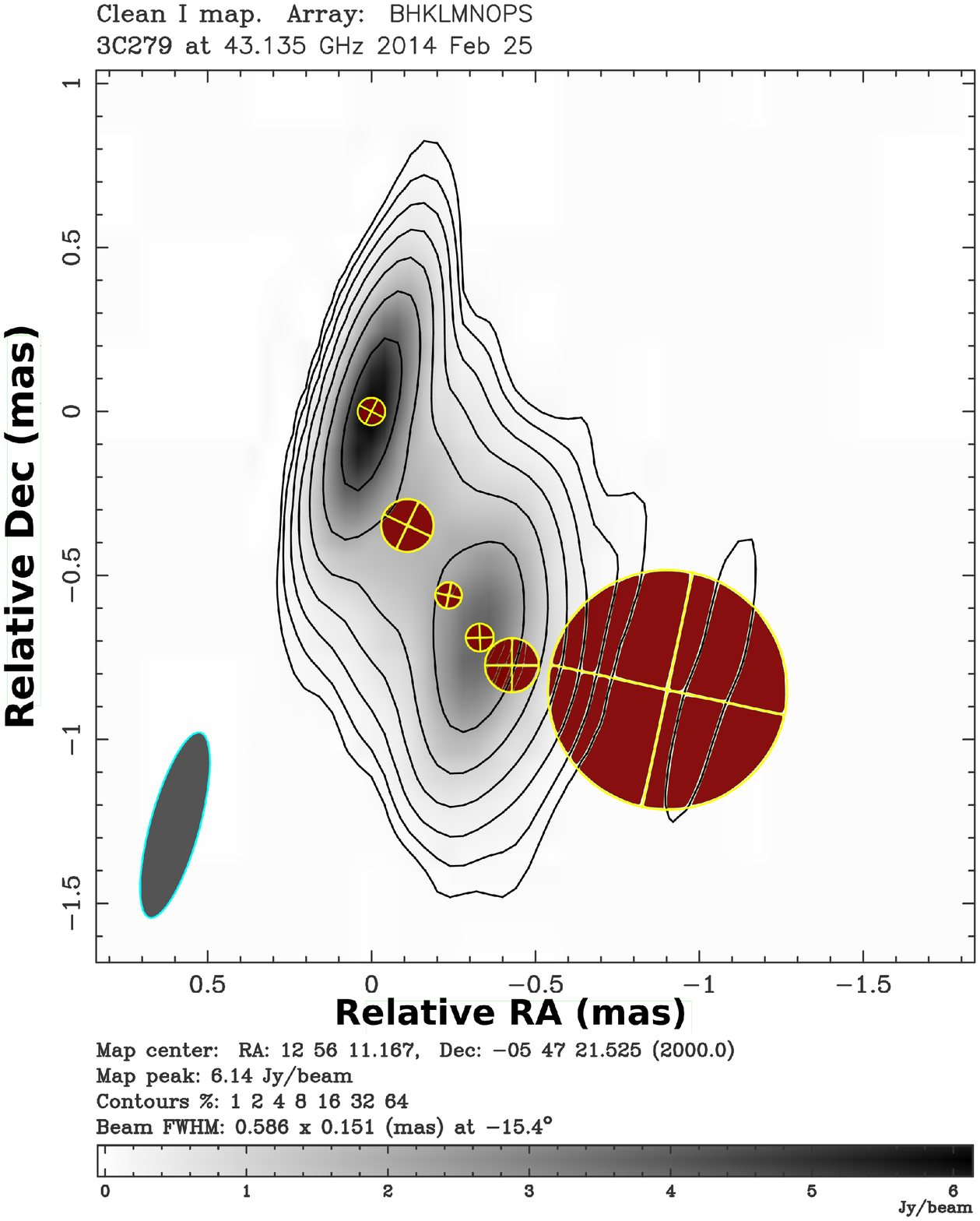}
\includegraphics[scale=0.44,angle=0, trim=0 0 0 45, clip]{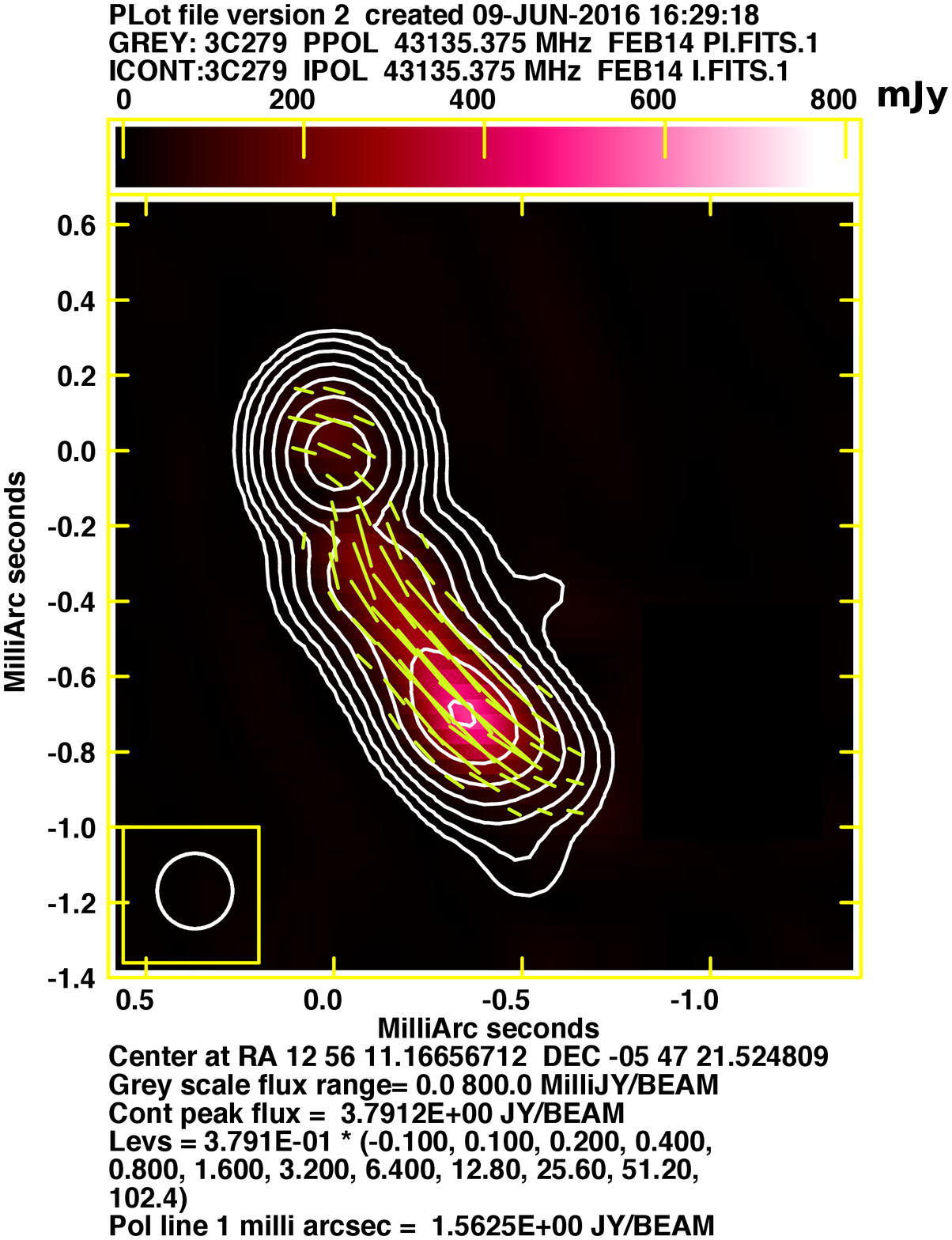}
  \caption{Top: An example image of 3C~279 as observed in February 2014: Gaussian circular 
components plotted on top of the total intensity contours. Bottom: Composite total (contours) and polarized (color) intensity image of 3C~279 convolved with a beam size of 0.1~mas. The line segments 
(length of the segments is proportional to fractional polarization)  show the EVPA direction. 
The ellipse/circle in the bottom left corner is the synthesized beam of the array.     }
\label{plot_fig7}
\end{figure}

In Fig.\ \ref{plot_fig7}, we display an example image of  the 3C~279 jet. 
The top panel shows the circular Gaussian components (yellow crossed-circles) 
superimposed on top of the total intensity contours. The bottom panel shows the composite 
super-resolved image convolved with a beam size of 0.1~mas (0.63~parsec). In the composite 
image, the contours represent the total intensity and the color scale represents the polarized 
intensity. The yellow line segments mark the EVPA direction. 
 
The total intensity of the jet is described using circular Gaussian 
components, and we list the model fit parameters  in Table \ref{tab1} in Appendix. To study 
the evolution of flux density and polarization of components, and also the component 
motion and their evolution as a function of time, we cross-identify the features 
using the following assumption. We assumed that the changes in the flux density, 
distance from the VLBI core, position angle, and size should be small for the time 
period between adjacent epochs to prevent a potentially large systematic error arising 
from the incorrect cross-identification.

\subsubsection{Component motion} 

\label{comp_motion}
\begin{figure*}
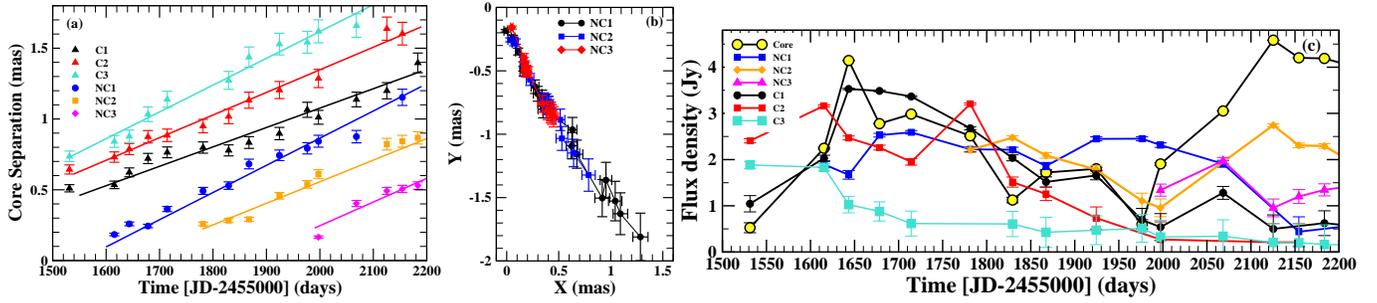

\includegraphics[scale=0.31,angle=0, trim=1.3 0 0 0.5, clip]{jet_dist.eps}
\includegraphics[scale=0.31,angle=0, trim=0 0 0 0.9, clip]{jet_dist_XY.eps}
\includegraphics[scale=0.38,angle=0, trim=0 0 0 0.2, clip]{jet_flx.eps}
  \caption{(a) Temporal evolution of the components' radial separation from the core; the solid 
lines represent fits with a linear function. (b) Trajectories of the components in the X-Y plane 
(see Section \ref{comp_motion} for details). (c) Flux density variations of the individual knots. }
\label{plot_fig8}
\end{figure*}

 The VLBI core  centered at (RA 12 56 11.167 DEC -05 47 21.53) is chosen as 
a reference point, and is fixed 
to coordinates (0, 0) to 
study the kinematics of individual components. The 
absolute position of the core could  change because of changes in opacity, pressure or instabilities in the jet \citep[see][for details]{hodgson2017, lisakov2017}.  
 However either we fix the core position to (0,0) or leave it free, it does not affect the relative separation of the components w.r.t. the core.
During the course of our campaign period, we identified a total of 6 
components - C1, C2, C3, NC1, NC2, and NC3  - in addition to the core. Figure \ref{plot_fig8} shows the evolution of the radial distance of the components from the core (a) and their trajectories 
in the XY-plane\footnote{X = $r$~cos($\theta$ + 90$^{\circ}$) 
and Y = $r$~sin($\theta$ + 90$^{\circ}$); $r$ is the radial separation from the core and $\theta$ is the position 
angle with respect to an imaginary north-south line drawn through the map center} projected on the sky (b). 
The flux density variations in the individual knots are shown in Fig.\ \ref{plot_fig8} (c).

In August 2013, the jet's brightness could be well described using five components 
(C1 to C3 and two unidentified components) in addition to 
the core  (see Tab.\ \ref{tab1} in Appendix). A new component (NC1) appeared  in November 
2013, followed by the ejection of two more components in May (NC2) and December (NC3) 2014. The core brightened before the ejection of a new component and 
later the component appeared in the jet. All these components exhibit significant motion especially in 
the radial direction. As seen in Fig.\ \ref{plot_fig8} (b), the components tend to follow the same 
linear trajectories in the XY-plane up to a distance of $\sim$0.5~mas. This 
is in contrast to what we usually observe for blazars. Wiggly/curved trajectories 
of components  are often  seen for blazars \citep[see][and references therein]
{rani2014, rani2015, chatterjee2008, lister2013}.

To study the component motion in the radial direction, we fitted the radial separation of 
the components relative to the core using, $r(t) = \mu \times (t - T_0)$, where $\mu$ is the angular 
speed in mas~yr$^{-1}$ and $T_0$ corresponds to the ejection time of the components from the core.  
The solid lines in Fig.\ \ref{plot_fig8} (a) represent the best-fitted functions.  The fit provides 
the angular speed ($\mu$), which we used to calculate the apparent velocity following 
$\beta_{app} = \frac{\mu.D_L}{c.(1+z)}$, where $D_L$ is the luminosity distance and $z$ is the 
redshift of the source. In Table \ref{tab2}, we list the angular and apparent speeds of the 
components. Back-extrapolation of the components' motion is used to estimate the ejection time 
of the new components (NC1 to NC3). The calculated ejection times  of the new components are 
listed in Table \ref{tab2}.  The highest detected apparent speed ($\beta_{app}$) can be used to 
get an estimate on the lowest possible Lorentz factor, $\Gamma \geq (1 + \beta_{app}^{2})^{0.5}$  
and maximum viewing angle, $\theta_{view} \leq sin^{-1}(1/\beta_{app})$. Using $\beta_{app}$ = 22.4, 
we get $\Gamma \geq$~22.4 and $\theta_{view} \leq$~2.6$^{\circ}$.

\begin{table}
\center
\caption{Kinematical parameters of the jet emission}
\begin{tabular}{l c c c  } \hline
Component & $\mu$ (mas~yr$^{-1}$)& $\beta_{app}$ (c)    & T$_0$ (days)          \\\hline 
C1  &0.49$\pm$ 0.04 &15.8$\pm$ 1.2  & --                    \\
C2  &0.58$\pm$ 0.03 &18.4$\pm$ 0.8  & --                   \\
C3  &0.69$\pm$ 0.02 &21.8$\pm$ 0.7  & --                   \\
\smallskip
NC1 &0.71$\pm$ 0.02 &22.4$\pm$ 0.7  & 1558$_{+15}^{-28}$       \\
\smallskip
NC2 &0.55$\pm$ 0.03 &17.5$\pm$ 0.8  & 1638$_{+40}^{-27}$    \\
\smallskip
NC3 &0.61$\pm$ 0.04 &19.2$\pm$ 1.2  & 1810$_{+26}^{-19}$    \\\hline
\end{tabular} \\
$\mu$ : angular speed \\ 
$\beta_{app}$ : apparent velocity \\ 
T$_0$ : component's core separation time \\
\label{tab2}
\end{table}

\subsection{Polarization variations}
\label{sect_pol_var}
\begin{figure}
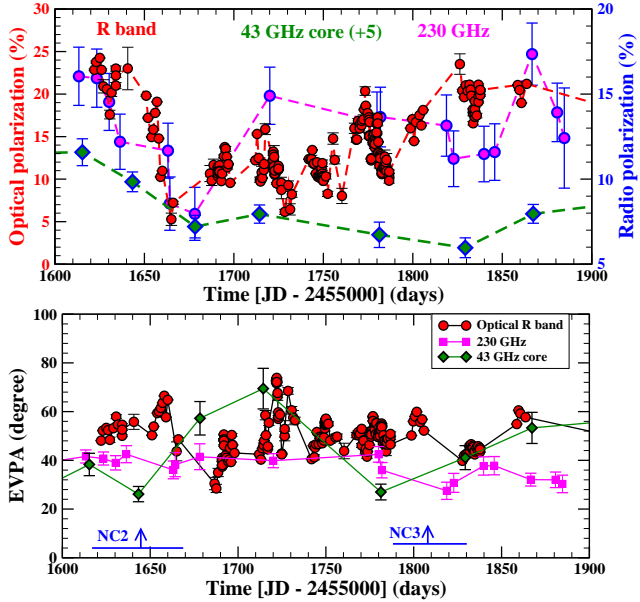

\includegraphics[scale=0.35,angle=0, trim=1 1 0 1, clip]{Radio_core_Opt_PP.eps}
\includegraphics[scale=0.35,angle=0, trim=1 1 0 1, clip]{Opt_radio_SD_core.eps}
\caption{{\it Top:} A comparison of the percentage polarization  variations 
at radio  and optical frequencies. For illustration purposes, the 43 GHz radio core 
percentage polarization curve is shifted  by $+$5~Jy (left scale should be used for the optical 
data and right for the radio data).  EVPA variations are displayed in the 
{\it bottom} panel.      }
\label{plot_fig11}
\end{figure}

Figure \ref{plot_fig11} presents a comparison of the variations observed in 
percentage polarization (PP) at optical and radio frequencies (left y-scale is for 
the optical data while the right is for the radio data). Although it is difficult to 
make a one-to-one comparison between the PP variations at the two frequencies because of 
sparse sampling of the radio data but we do see a similarity. For instance, in the beginning 
the PP is high both at optical and radio bands. A drop in PP (between JD$^{\prime}$ 1600 and 1680) is 
observed almost simultaneously  at the two frequencies. An increase in PP appears after the dip  
observed in optical and 230 GHz data  but not in the 43~GHz VLBI data. PP in the core region 
remains rather constant after the dip. On the other hand, PP at optical and 230 GHz radio frequencies 
continue rising until  they regain their initial value  at the end of the observations.

Figure \ref{plot_fig11} (bottom) shows a comparison of the EVPA variations at optical and 
radio frequencies. In contrast to PP, EVPA variations are rather modest.
A peak to peak (minimum to maximum) change of $\sim$60$^{\circ}$ is observed 
in the optical and 43 GHz VLBI data, and in the 230 GHz data, it is $\leq$25$^{\circ}$. 
The average EVPA values at the three frequencies are also quite 
similar ($\sim$40$^{\circ}$ -- 60$^{\circ}$). 
This indicates that the optical and 
radio polarization angles (EVPAs) are roughly parallel to the jet 
direction\footnote{Since at optical frequencies, we are already in the 
optically thin regime, the Faraday  effects are negligible here.} 
($\sim$$-$135$^{\circ}$).  We noticed a change  in the VLBI core EVPA during the ejection of 
new components (NC2 and NC3), which could be because of a different EVPA orientation of the new 
component(s).

\begin{figure}
\includegraphics[scale=0.35,angle=0, trim=1 1 0 1, clip]{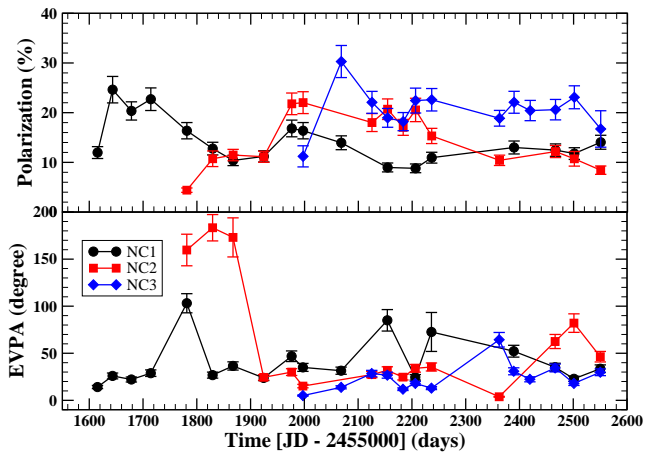}
\caption{Percentage polarization and EVPA  variations observed in the individual components.   }
\label{plot_fig12}
\end{figure}

The PP (top) and EVPA (bottom) variations observed in the new ejected 
components are shown in Fig.\ \ref{plot_fig12}.   
The PP of the components gradually increase after their
separation from the core and stay as high as $\geq 20\%$. This
could result from either a transition from the optically
thick to thin regime at 43~GHz or a progressive ordering of the
magnetic field with distance down the jet.
The EVPAs of the new components (especially NC2) show some modest variations 
but still remain aligned along the jet axis.

  \begin{figure*}
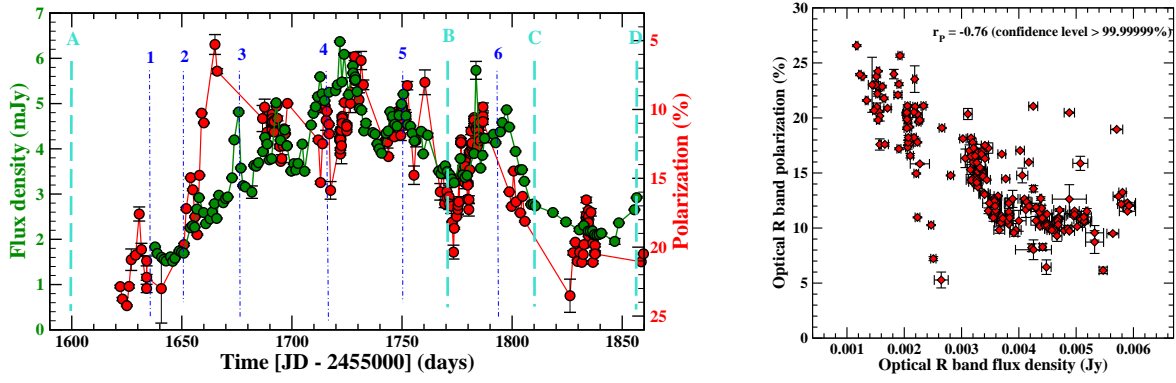

\includegraphics[scale=0.37,angle=0, trim=0 0 0 0, clip]{Opt_flx_pol.eps}
\hspace{0.9cm}
\includegraphics[scale=0.3,angle=0, trim=0 0 0 0, clip]{opt_flx_pol_linear_corr.eps}
\caption{{\it Left:} Optical R band flux density variations superimposed with the optical percentage polarization variations (vertical dotted and dashed lines are same as in Fig.\ 1). The right y-scale is for optical polarization, 
and  is reversed   for comparison purposes.   {\it Right:} Optical polarization 
vs.\ flux density  plot.     }
\label{plot_fig13}
\end{figure*}

  \begin{figure*}
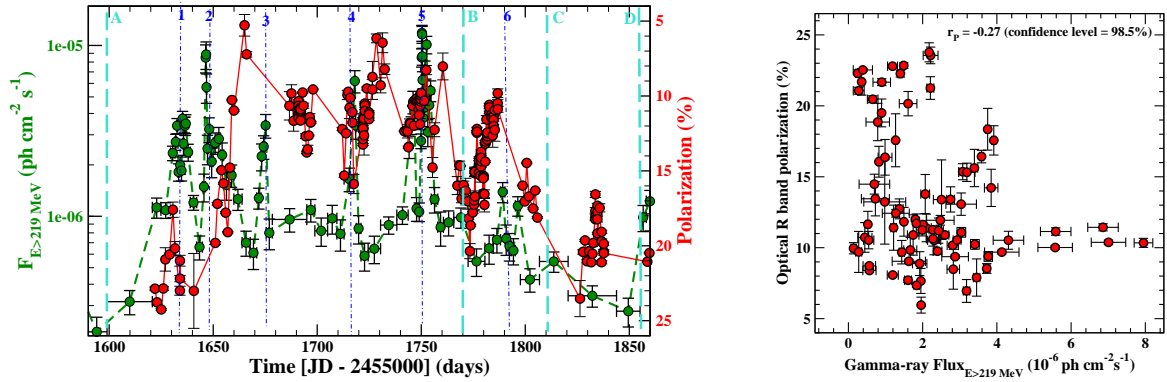

\includegraphics[scale=0.35,angle=0, trim=0 0 0 0, clip]{Opt_pol_gamma.eps}
\hspace{0.9cm}
\includegraphics[scale=0.29,angle=0, trim=0 0 0 0, clip]{gamma_opt_pol_linear_corr.eps}
\caption{{\it Left:} Gamma-ray photon flux variations superimposed on the optical percentage polarization 
variations (vertical dotted and dashed lines are same as in Fig.\ 1). The right y-scale is for optical polarization, 
and  is reversed   for comparison purposes.   {\it Right:} Optical polarization 
vs.\ $\gamma$-ray photon flux plot.     }
\label{plot_fig14}
\end{figure*}

\subsection{Flux density vs.\ polarization variations} 
\label{corr_flx_pol}
In Fig.\ \ref{plot_fig13}, we show a comparison of optical flux density 
and polarization variations. We used an inverse y-scale for the optical 
PP data. During the first 3 $\gamma$-ray flares 
(``1" to ``3"), the optical flux is rising and a decay is seen in the 
PP. Peaks of flares ``4" to ``5" coincide with the dips in the PP curve 
indicating a clear anti-correlation between the two. The anti-correlation 
is even more evident in the PP vs.\ optical flux plot  on the right. 
A formal linear Pearson correlation test suggests a significant anti-correlation 
between the optical flux and polarization data; we obtained $r_P$ = $-$0.76, at the
$>$99.99999$\%$ confidence level.

  \begin{figure}
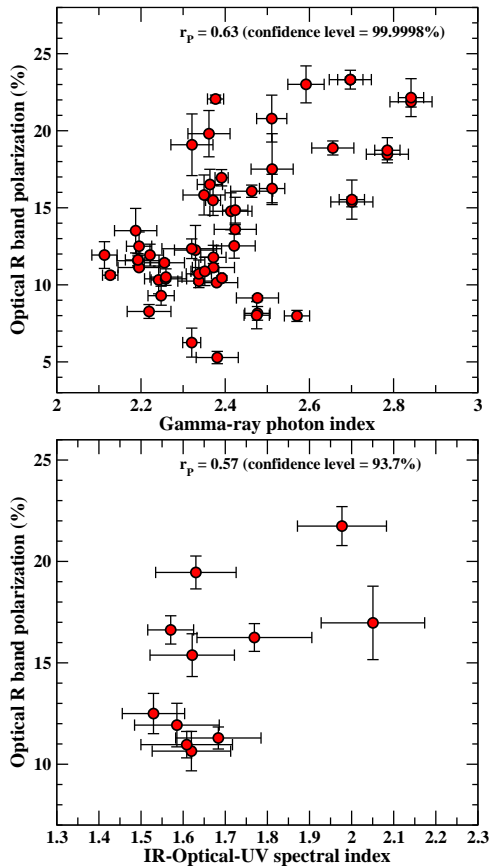

\includegraphics[scale=0.35,angle=0, trim=0 0 0 0, clip]{gamma_indx_opt_pol_linear_corr.eps}
\includegraphics[scale=0.35,angle=0, trim=0 0 0 0, clip]{Opt_indx_pol_linear_corr.eps}
\caption{ Optical polarization vs.\ $\gamma$-ray photon index ({\it top}) and IR-optical-UV 
spectral index ({\it bottom}).     }
\label{plot_fig15}
\end{figure}

  \begin{figure*}
\vspace{-1.9cm}
\includegraphics[scale=0.48,angle=0, trim=0 0 0 1.2, clip]{frac_Q_U.eps}
\hspace{0.9cm}
\includegraphics[scale=0.25,angle=0, trim=0 0 0 8, clip]{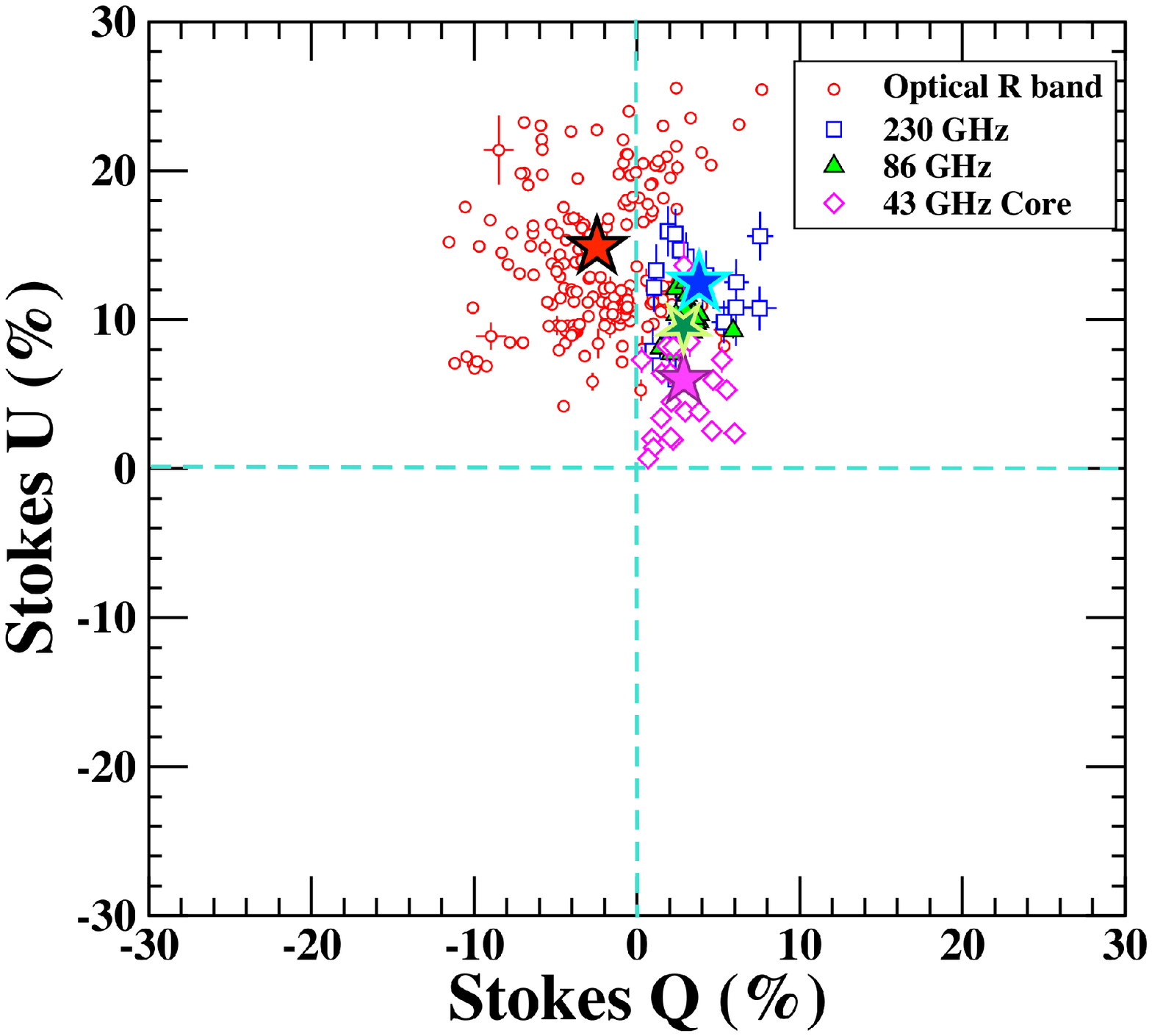}
\caption{{\it Left:} Optical R band $Q_{frac}$ and $U_{frac}$ variations (vertical dotted and dashed 
lines are same as in Fig.\ 1). {\it Right:}  Stokes $Q_{frac}$ versus $U_{frac}$; centroid 
in each case is marked with a star.      }
\label{plot_fig16}
\end{figure*}

A comparison of the optical PP and $\gamma$-ray photon flux variations is shown in 
Fig.\ \ref{plot_fig14} (left) (again an inverse-scale is used for the optical PP data). 
There are hints of an apparent anti-correlation between the two.  For instance, at the onset 
of the $\gamma$-ray outburst (JD$^{\prime}$ $\sim$1610), the optical PP is quite high 
($\sim$25$\%$).  Also, the peaks of $\gamma$-ray flares (``4" and ``5") roughly coincide 
with the dips in the PP curve. Finally, the optical PP 
regained its original value ($\sim$25$\%$) at the end of the outburst (JD$^{\prime}$ $\sim$1820). 
 The anti-correlation is even more evident in the optical PP vs.\ $\gamma$-ray 
photon flux plot shown in the right panel of Fig.\ \ref{plot_fig14}. The statistical significance 
is tested via the linear-Pearson correlation analysis, which suggests a significant 
($r_P$ = $-$0.27 at a confidence level = 98.5$\%$) anti-correlation between the two.

Figure \ref{plot_fig15} shows the PP vs.\ spectral index plot. We again used the linear-Pearson 
correlation analysis to test the significance of the correlation 
between PP and optical/$\gamma$-ray  spectral indices.
Formally we obtained $r_P$ = 0.63 at the 99.9998$\%$ 
confidence level for the PP vs.\ $\gamma$-ray photon index data (see top panel), which suggests 
a significant correlation between the two. The PP vs.\ optical spectral index plot suggests 
a correlation ($r_P$=0.57) at the 93.7$\%$ confidence level. 
 A positive correlation between PP and optical/$\gamma$-ray spectral indices 
indicates that  higher PP is observed for a steeper spectrum. A significant 
hardening of the optical/$\gamma$-ray spectrum is observed during bright flares (see Section \ref{lc_gamma} and \ref{lc_opt} for details).

Change in PP could be simply explained either by a change in B-field order (i.e.\ an increase 
in PP indicates a more ordered magnetic field)  or by a change 
in the spectral index of the emitting particles.   We could  expect to see some 
increase in PP due to 
steepening of the electron spectrum or the synchrotron spectrum (degree of maximum polarization = 
($\alpha$ + 1)/($\alpha$ + 5/3), where $\alpha$ is the spectral index). For instance, steepening of a spectrum 
by one magnitude  ($\Delta \alpha =1$) produces a $\leq$5$\%$ increment in PP magnitude. 
Although the PP significantly correlates  with $\alpha$, our observations suggest 
variations of $20\%$ ($\Delta$PP =20). This in turn suggests that the PP variations   cannot be solely explained via steepening of  the  spectrum as a change in the spectral index 
has only a very slight affect on the PP.

Since the mean polarization is the maximum divided by the square-root of the number of 
cells\footnote{the smallest scale emission region having a preferred magnetic field 
orientation \citep{marscher2014}} 
involved in the emission, the observed anti-correlation between gamma-ray flux and optical PP 
could be because of the presence of  more turbulent cells during the outburst than at other times. 
The jet plasma is highly dynamic; plasma instabilities \citep[e.g., the current-driven kink instability and/or magnetohydrodynamical instabilities][]{nishikawa2016, nalewajko2012} could occur and vary with time, which basically drives the 
turbulence. In short, instabilities drive turbulence which can influence the number of cells in a given 
emission region.    This in turn implies that the 
turbulence could enhance the efficiency of particle acceleration, 
which is why there is an 
outburst and also why the spectrum is flatter during the outburst and steeper before and 
after the outburst when the PP is higher. 
An observational test of the turbulence cell model \citep[TEMZ,][]{marscher2014} 
is to compare the standard 
deviation of random PP fluctuations (compared to the systematic variations). In case of the presence of only turbulent component(s) $\sigma$(PP) = 0.5$<$PP$>$. 
We used equal time bin when the PP is not changing systematically to 
determine $\sigma$(PP).
For 3C~279 the calculated  $\sigma$(PP) = (0.08 to 0.25)~$<$PP$>$, which suggests the presence of a polarized steady component as well or the presence of multiple emission 
components with different polarizations.

To further test the hypothesis of the presence of a steady polarized component in 
addition to the multiple turbulent components, we investigate the variations in Stokes Q and U 
parameters. Photometric and polarimetric 
measurements obtained on the same day were used to calculate the fractional Stokes parameters, 
$Q_{frac}$ = PP~cos(2$\times$EVPA) and $U_{frac}$ = PP~sin(2$\times$EVPA). Figure \ref{plot_fig16} 
(left) shows the variations in optical R band $Q_{frac}$ and $U_{frac}$. A clear difference 
in the Stokes $Q_{frac}$ and $U_{frac}$ variations can be noticed here.
The observed  dependence between the Stokes parameters is given in the right panel of 
Fig.\ \ref{plot_fig16} (red: optical R band, magenta: 43~GHz radio core, blue: 230 GHz, and 
green: 86 GHz).  The $Q_{frac}$, $U_{frac}$ centroid clearly deviates from  (0,0). We noticed 
similar shift at 86 and 230 GHz radio bands, and also in  the 43 GHz radio core.   
The observed shift of the Q,U centroid  with respect to (0,0)  can be explained by the presence 
of a steady components with  polarization direction aligned along the jet axis \citep{blinov2009}. 
Moreover, a similarity of EVPA values at optical and radio frequencies with the VLBI core EVPAs 
supports the alignment of optical/radio polarization direction along the jet axis 
(Section \ref{sect_pol_var}). As is suggested by \citet{blinov2009}, if the steady polarized 
component has a higher degree of polarization, one can expect to see an anti-correlation 
between the flux and degree of  polarization.  The observed PP  at the beginning/end 
of the outburst (see Figs.\ \ref{plot_fig13} and \ref{plot_fig14}) is $\sim$25$\%$, which 
means that the steady polarized component is higher than 25$\%$.   A steady 
polarized component having its direction parallel to the jet axis could 
be  the toroidal magnetic field component of the jet.  Alternatively,   
the presence of multiple emission components with different polarizations could be 
plausible as well.

\subsection{Connection between jet activity and broadband emission}
Given the superposition of multiple modes of flaring activity, it is quite challenging 
to understand the broadband flaring behavior of the source. The major outburst (A to D) 
is accompanied by
faster flares across the whole electromagnetic spectrum (see Fig.\ \ref{plot_lcs}). 
Interestingly, a new component ``NC2" is 
ejected from the core during the period of the first three (``1" to ``3") $\gamma$-ray flares. 
The next three flares (``4" to ``6") were followed by a 
continuous decay in the source brightness 
at $\gamma$-ray, optical, and radio frequencies  between period ``C" and ``D"). Later, the ejection of a new 
component ``NC3" from the core suggests a delay between the high-energy activity  
and the component's core-separation timing.

 Simulations (as described in Appendix \ref{simulations}) were used to test whether 
the connection between component ejection and $\gamma$-ray flaring activity is 
just by chance or not. First we simulated a total of 1000 light curves. Next, 
we checked  how often do we see a gamma-ray flare  in the simulated data during the ejection of 
NC2 (JD$^{\prime}$ = 1611 to 1678) and NC3 
(JD$^{\prime}$ = 1790 to 1836). Given the superposition of multiple modes of flaring activity, 
defining a flare could be tricky. We choose a simplistic approach here. As the 
typical duration of the rapid $\gamma$-ray flares (1 to 6) is 10 to 30~days, we define 
a continuous rise in the $\gamma$-ray flux above 6.6$\times$10$^{-7}$~ph~cm$^{-2}$~s$^{-1}$ (two times above the quiescent level) for $\geq$20~days as a flare, and tested 
how often this condition is met in our simulated light curves. 
During the NC2 ejection period, we found a flare in 28 out of 1000 cases; in short, the ejection 
of NC2 is  connected to the $\gamma$-ray flaring activity with  97.2$\%$  confidence level. Similar, we 
found that the $\gamma$-ray flares are related to NC3 with a 95.8$\%$ confidence level. 
We also tested the significance of connection between $\gamma$-ray 
flares observed within 50~days before and after the ejection of the components. Our calculations 
suggest that the chance-coincidence of NC2 being associated to a flare(s) 50~days before or 
after its ejection is below 90$\%$. In case of NC3, there are fair chances (94.2$\%$) that its ejection 
could be related to a flare(s) observed 50 days prior to the ejection, while the confidence-level of its 
ejection being connected to a flare observed with 50 days after to its  ejection is below 90$\%$.  
Moreover, we tested the chance-coincidence for the different flare 
durations. For a short flare duration of 10~days, the probability of observing a flare by chance 
during the ejection of NC2 (and NC3) is less than 10$\%$ i.e.\ the confidence level of NC2 (and NC3) 
being associated with a $\gamma$-ray flare is $\simeq$90$\%$. In case of a long-duration flare (30~days), 
the confidence level is $>$96$\%$.

\subsection{Multiple energy dissipation regions}   
During 2013-2014, we observed two modes of flaring activity (faster flares sitting 
on top of the long-term outburst) in 3C~279 at radio, optical and $\gamma$-ray energies. 
Although X-ray counterparts were observed for two flares, the long-term outburst seems 
to be missing. The broadband SED  modeling during this 
period \citep[reported in][]{hayashida2015}  suggests that single zone emission 
models that  matched the optical and $\gamma$-ray spectra failed to describe the 
X-ray emission, which suggests the presence of multiple emission regions.

Unlike the last three  flares (``4" to ``6"), the first three 
$\gamma$-ray flares (``1" to ``3") are missing their  optical/radio counterparts, 
except for flare ``3" which is observed at optical bands. Moreover, the 
IOU spectrum is relatively softer ($\alpha_{IOU}$ $\sim$1.8--2.0) during the 
first episode of flares.  Later, the IOU spectrum gets 
harder ($\alpha_{IOU}$ $\sim$1.6). 
A new component, NC2, appeared in the jet during the  flare 
``1" to ``3".  
Ejection of NC2 during the first three flares  indicates that 
these high-energy flares have to be produced either at the VLBI core or very 
close to the core region; 
otherwise, we would have observed a delay between the high-energy flares and the 
component ejection, which we did not. A delay is observed between the last three 
flares (4 to 6) and the ejection of NC3. This indicates that the high-energy region is 
located further upstream of the VLBI core (closer  to the central BH). 
Moreover, the PP vs. $\gamma$-ray/optical flux and spectral index correlation 
(discussed in Section \ref{corr_flx_pol}) suggests the presence of multiple 
turbulent cells in a given emission region and also that the number 
of these cells is variable with time.

\begin{figure*}
\includegraphics[scale=0.1,angle=0, trim=3 0 13 0, clip]{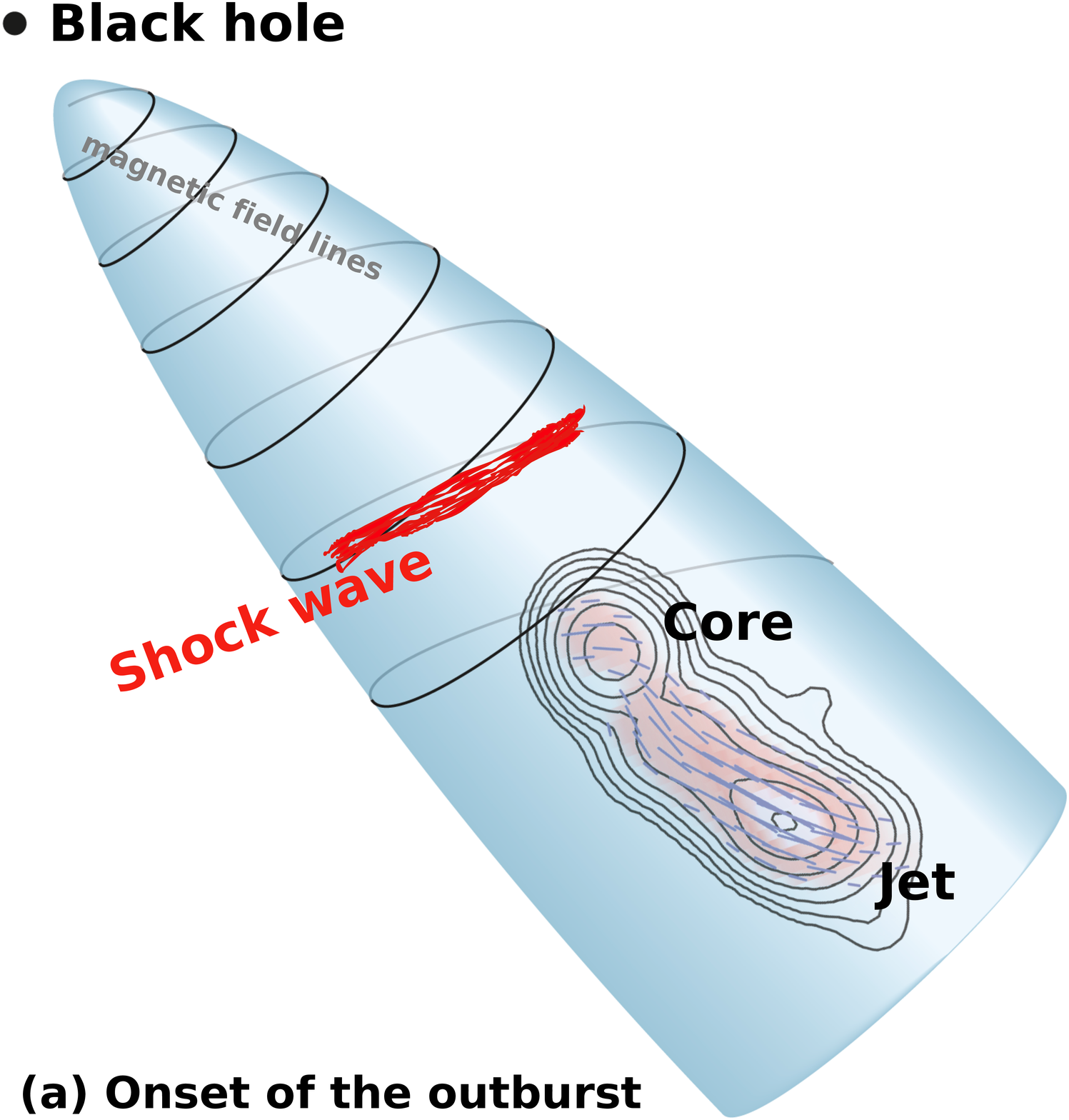}
\includegraphics[scale=0.1,angle=0, trim=3 0 13 0, clip]{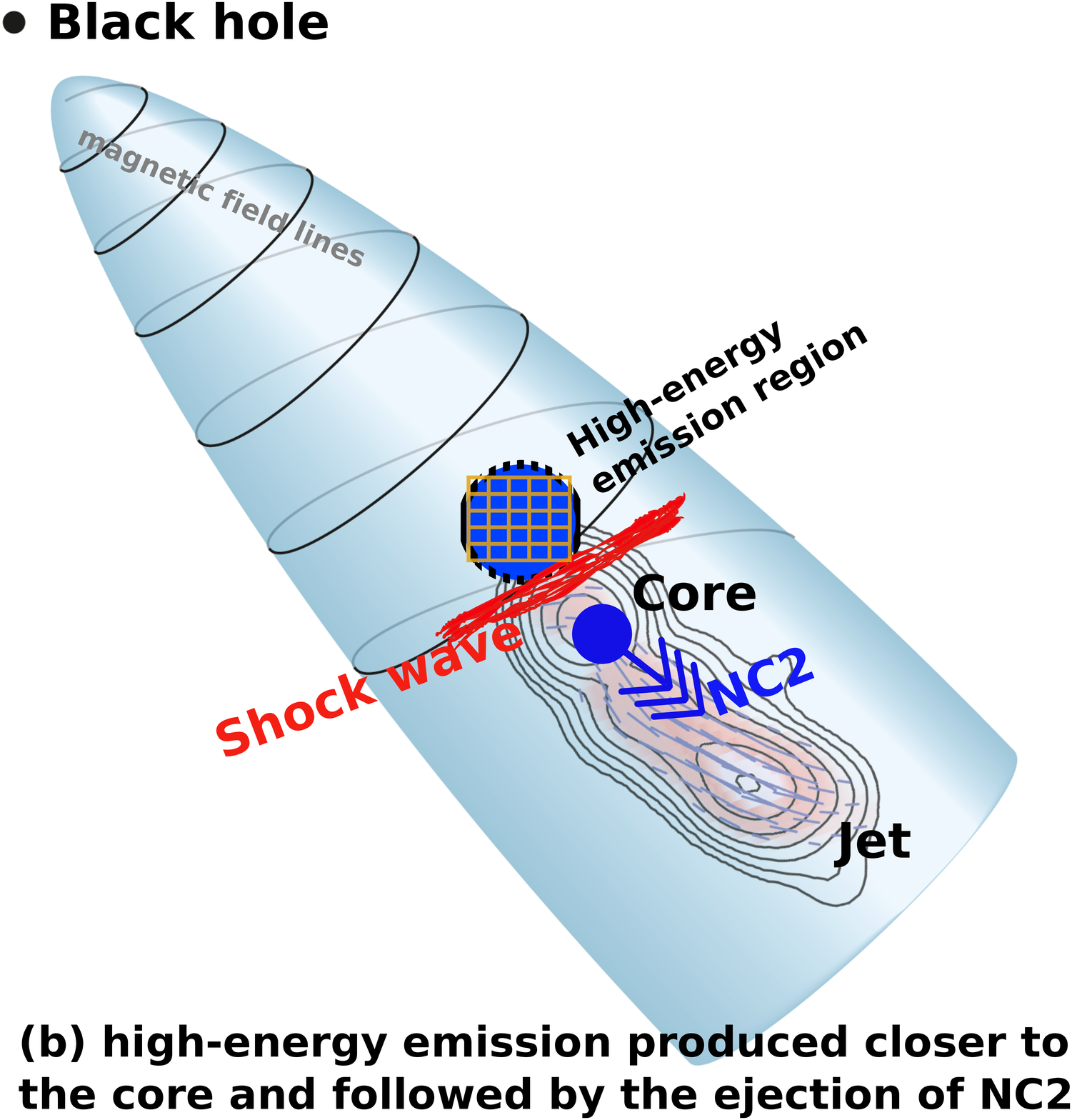}
\includegraphics[scale=0.1,angle=0, trim=3 0 13 0, clip]{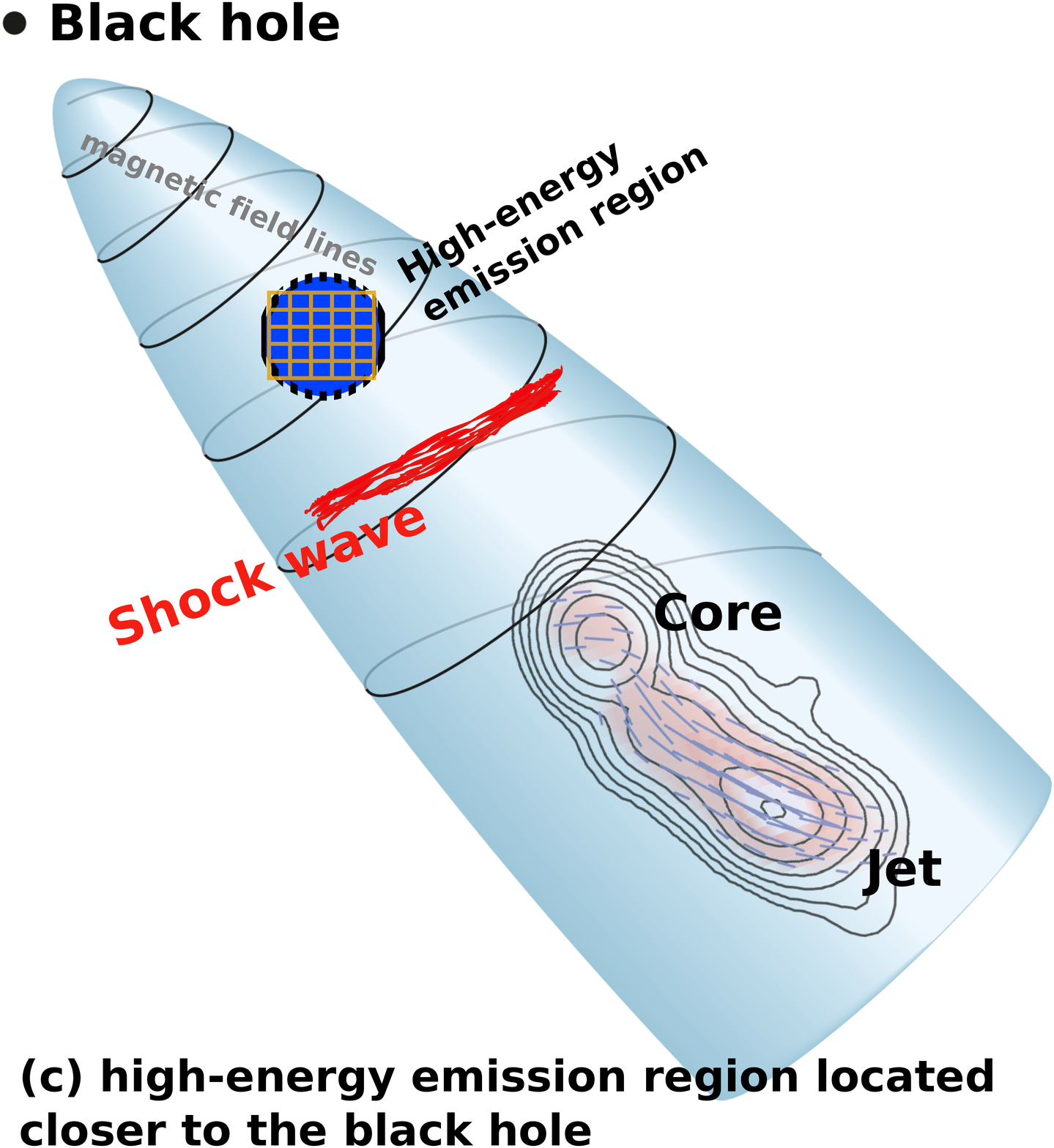}
\includegraphics[scale=0.1,angle=0, trim=3 0 13 0, clip]{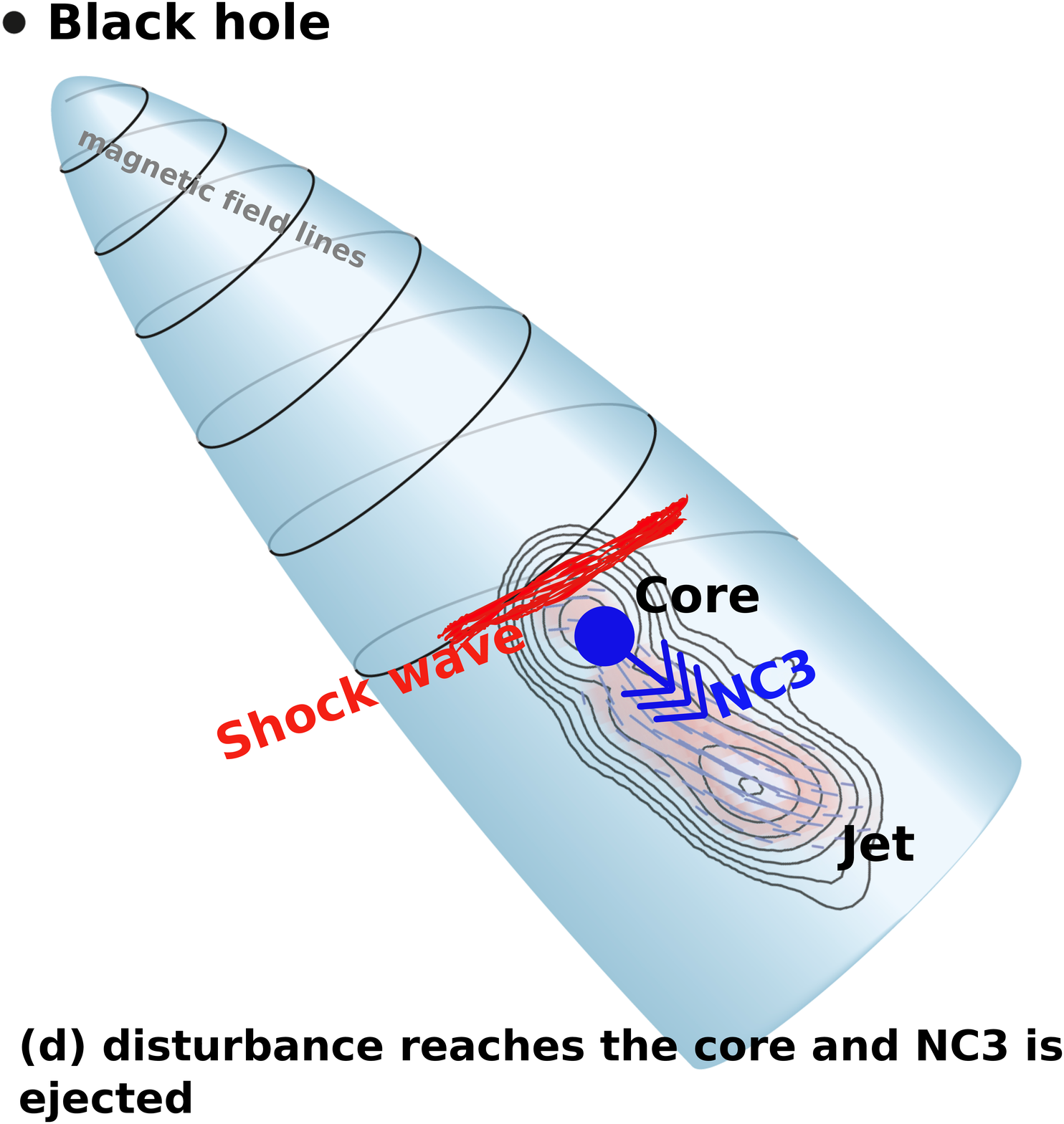}
\caption{Sketch for the proposed scenario in the 3C~279 (not to scale): a disturbance 
(or shock wave) reaches the core(a) producing the high-energy flares (``1" to ``3") closer to the core 
region and an ejection of component NC2 (b).  (c)  High-energy emission region is located closer to the 
black hole, and later, the disturbance 
passes the core resulting into the ejection of NC3 (d).     }
\label{plot_fig17}
\end{figure*}

 We interpret the observed broadband flaring activity in the following 
manner (see Fig.~\ref{plot_fig17}). A disturbance at the base of the jet could 
cause a shock wave and propagate down the jet, and could be
related  to the onset of the major outburst (Fig.~\ref{plot_fig17} a). Later, 
the moving shock  
reaches  the 43~GHz core producing the first 3 flares 
(1 to 3). Given the 
different B-field properties of cells, the particle acceleration efficiency 
could be different for different cells. Since at a given frequency, 
what we observe is their collective emission,  concurrent broadband flaring 
activity may or may not 
be observed. The moving shock disturbs the jet outflow and becomes knot NC2 as it 
separates from the core (Fig.~\ref{plot_fig17} b). 
In the second episode of flaring activity (4 to 5), the emission region 
seems to be located further 
upstream of the core (closer to the central black hole, see: Fig.~\ref{plot_fig17} c). Later, the disturbance/shock wave passes the 
core region  resulting in the ejection of NC3 (Fig.~\ref{plot_fig17} d). Our analysis therefore supports two possible 
energy dissipation sites in 3C 279. 
Location of the dissipation zone depends on the parameters of a disturbance, 
especially the Lorentz factor ($\Gamma$), and for disturbances with lower $\Gamma$ the dissipation zone
is expected to be closer to the BH than for those with larger $\Gamma$ 
\citep{katarzynski2007}.
Although $\beta_{app}$ of NC2 and NC3 are similar within the uncertainties,
it is possible that $\Gamma_{NC2}$ $>$ $\Gamma_{NC3}$ that would result in the dissipation zone during the flares 1-3 
to be farther from the BH and closer to the VLBI core.
Both observations \citep{rani2015, hodgson2017, jorstad2005} 
and simulations \citep{gomez1997, fromm2012} support the presence of standing shocks in parsec-scale AGN jets. In fact 
at short mm-radio bands, the VLBI core could be a standing shock as well \citep{hodgson2017, marscher2008}. Passage of a 
moving shock through standing shocks is quite efficient in accelerating particles to high energies, which could explain  
why some sites are more favored than others for energy dissipation \citep{gomez1997, meyer2015}.

\section{Summary and conclusion} 
We used high-frequency VLBI observations to probe the broadband flaring activity in 
3C 279. At GeV energies the source went through a series of flares superimposed on a 
long-term outburst from JD$^{\prime}$ = 1600 to 1850 (November 2013 to August 2014). In total we observed 6 $\gamma$-ray 
flares, and four of these flares have optical (and radio) counterparts. 
X-ray counterparts are only observed for two flares (2 and 5). Moreover, similar to $\gamma$ rays, 
we observed two different modes  of flaring activity (fast flares superimposed on a long-term 
outburst) at optical and radio frequencies. On the other hand, the long-term outburst activity 
seems to be missing at X-rays.

Two modes (fast flares superimposed on long-term variations) of variability are also observed 
in percentage polarization (PP) data at optical and radio frequencies.  On the other hand, the polarization angle 
(EVPA) variations are quite modest at both optical and radio frequencies, and the EVPAs are 
well aligned with the jet direction.

Two new components (NC2 and NC3) were ejected from the core during the 
flaring activity  period, and are connected to the source flaring activity 
with $\geq$95$\%$ probability.  NC2 appeared in the jet during the 
first three flares (1 to 3)  pointing to the  
the 43~GHz VLBI core as the possible location of energy dissipation. Ejection of NC3 happened  
after the last three flares (4 to 6), which puts the location of energy dissipation further 
upstream of the VLBI core (in a region closer to the BH). Our analysis therefore  argues in favor  
of multiple energy dissipation regions in the source.

We detected an anti-correlation between the optical PP and $\gamma$-ray/optical 
flux variations, while, a positive correlation is seen for the PP and $\gamma$-ray/optical 
spectral index variations. We propose that this could be explained  using the turbulent  
multi-zone cell model (TEMZ). If there are more turbulent cells in the emission region 
during the outburst than other times,  one might   
expect to see an anti-correlation between $\gamma$-ray/optical flux and PP. 
This in turn implies that the turbulence is related to the efficiency of particle 
acceleration.    The observations however solely cannot be described 
using the turbulent component. A shift in Stokes (Q,U) centroid w.r.t. (0,0)  
suggests the presence of  multiple emission components with different 
polarizations.

\acknowledgements
The \textit{Fermi}/LAT Collaboration acknowledges generous ongoing support
from a number of agencies and institutes that have supported both the
development and the operation of the LAT as well as scientific data analysis.
These include the National Aeronautics and Space Administration and the
Department of Energy in the United States, the Commissariat \`a l'Energie Atomique
and the Centre National de la Recherche Scientifique / Institut National de Physique
Nucl\'eaire et de Physique des Particules in France, the Agenzia Spaziale Italiana
and the Istituto Nazionale di Fisica Nucleare in Italy, the Ministry of Education,
Culture, Sports, Science and Technology (MEXT), High Energy Accelerator Research
Organization (KEK) and Japan Aerospace Exploration Agency (JAXA) in Japan, and
the K.~A.~Wallenberg Foundation, the Swedish Research Council and the
Swedish National Space Board in Sweden.
Additional support for science analysis during the operations phase is gratefully
acknowledged from the Istituto Nazionale di Astrofisica in Italy and the Centre
National d'\'Etudes Spatiales in France. This work performed in part under DOE
Contract DE-AC02-76SF00515.

This research was supported by an appointment to the NASA Postdoctoral Program
at the Goddard Space Flight Center, administered by Universities Space Research Association 
through a contract with NASA.
This study makes use of 43 GHz VLBA data from the
VLBA-BU Blazar Monitoring Program (VLBA-BU-BLAZAR; http://www.bu.edu/blazars/VLBAproject.html), funded
by NASA through the Fermi Guest Investigator Program. 
The VLBA is an instrument of the Long Baseline Observatory. The Long Baseline Observatory 
is a facility of the National Science Foundation operated by Associated Universities, Inc. 
The BU group acknowledges support from US National Science Foundation grant AST-1615796.
The Submillimeter Array is a joint project between
the Smithsonian Astrophysical Observatory and the Academia Sinica Institute of Astronomy and Astrophysics 
and is funded by the Smithsonian Institution and the Academia Sinica. 
This research was supported by an appointment to the NASA Postdoctoral
Program at the Goddard Space Flight Center, administered by
Universities Space Research Association through a contract with NASA. 
KS is supported by the Russian Science Foundation grant 16-12-10481.
The St.~Petersburg University team acknowledges support from Russian Science
Foundation grant 17-12-01029. 
This research has made use of data from the OVRO 40-m monitoring program 
\citep{richards2011} which is supported in part by NASA grants NNX08AW31G, 
NNX11A043G, and NNX14AQ89G and NSF grants AST-0808050 and AST-1109911.
IA acknowledges support by a Ram\'on y Cajal grant of the Ministerio de Econom\'ia, Industria, y Competitividad (MINECO) of Spain. Acquisition and reduction of the POLAMI data was supported in part by MINECO through grants AYA2010-14844, AYA2013-40825-P, and AYA2016-80889-P, and by the Regional Government of Andaluc\'ia through grant P09-FQM-4784. The POLAMI observations were carried out at the IRAM 30m Telescope. IRAM is supported by INSU/CNRS (France), MPG (Germany) and IGN (Spain).
BR acknowledges the help of  Roopesh Ojha, Marcello Giroletti, 
and Dave Thompson for fruitful discussions 
and comments that improved the manuscript.

\begin{appendix}
\label{appendix}
\section{Testing the significance of component ejections and $\gamma$-ray flares} 
\label{simulations}
The burst-like $\gamma$-ray flares in 3C~279 do not follow a Gaussian 
distribution and can be better described via log-normal or gamma distribution; therefore, we simulated the light curves using the Emmanoulopoulos \citep{emmanoulopoulos2013} 
method implemented by \citet{connolly2015}. 
The first step is to determine the 
power spectral density (PSD) slope of the observed $\gamma$-ray light curve. Given that the adaptive binned $\gamma$-ray data is unevenly sampled, we interpolated it using the cubic spline method in R\footnote{https://stat.ethz.ch/R-manual/R-devel/library/stats/html/splinefun.html}. 
As the data is manipulated via interpolation, we tested our PSD result using simulations. We refer to \citet{chidiac2016} for a step-by-step analysis.     
We found that the observed source variability can be well described by a PSD slope 
of $-(0.97\pm0.17)$. 
The method also takes into account of the underlying 
probability density function (PDF) of the given light curve.  
We simulated a 
total of 1000 light curves using the best-fit PDF and PSD parameters. 
The simulated light curves  are sampled at the same cadence as the 
adaptive binned $\gamma$-ray light curve. The average flux and the standard 
deviation of simulated data is equal to that  of the observed data.

\section{Model fit component results }
Table \ref{tab1} lists the model fit parameters of the 3C~279 jet. The jet kinematic study is 
presented in Section \ref{sect_vlbi}.  
\vspace{-0.5in}
\begin{table}
\begin{center}
\scriptsize
\caption{Results from Gaussian model fitting and component parameters }
\begin{tabular}{c c c c c c }\hline
Epoch  (JD$^{\prime}$)       & S        & r               & $\theta$           & $\phi$  & Comp$^{a}$ \\ 
 date & (Jy)         &(mas)            & ($^{\circ}$)       & (mas)   &    \\\hline
1531 & 4.52$\pm$0.45  &    0$\pm$0    &     0$\pm$0  & 0.054$\pm$0.005 & core   \\
(Aug.~2013)& 0.52$\pm$0.05  & 0.28$\pm$0.02 &  -173$\pm$17 & 0.092$\pm$0.009 & X      \\
     & 1.04$\pm$0.10  & 0.51$\pm$0.05 &  -160$\pm$16 & 0.074$\pm$0.007 & C1 \\
     & 2.40$\pm$0.34  & 0.64$\pm$0.06 &  -155$\pm$18 & 0.087$\pm$0.008 & C2  \\
     & 1.88$\pm$0.18  & 0.73$\pm$0.07 &  -153$\pm$15 & 0.154$\pm$0.015 & C3  \\
     & 0.38$\pm$0.03  & 1.13$\pm$0.11 &  -128$\pm$17 & 0.693$\pm$0.069 & X      \\
1615 & 6.24$\pm$0.62  &    0$\pm$0    &     0$\pm$0  & 0.043$\pm$0.004 & core    \\
(Nov.~2013)& 1.89$\pm$0.18  & 0.18$\pm$0.01 &  177$\pm$17 & 0.109$\pm$0.010 & NC1     \\
     & 2.02$\pm$0.30  & 0.53$\pm$0.05 &  -159$\pm$15 & 0.136$\pm$0.013 & C1  \\
     & 3.16$\pm$0.31  & 0.73$\pm$0.07 &  -156$\pm$20 & 0.114$\pm$0.011 & C2   \\
     & 1.83$\pm$0.18  & 0.84$\pm$0.08 &  -152$\pm$15 & 0.164$\pm$0.016 & C3   \\
     & 0.90$\pm$0.13  & 1.03$\pm$0.10 &  -137$\pm$22 & 0.753$\pm$0.075 & X       \\
1643 & 8.14$\pm$0.81  &    0$\pm$0    &     0$\pm$0  & 0.062$\pm$0.006 & core    \\
(Dec.~2013)& 1.66$\pm$0.16  & 0.25$\pm$0.02 &  -166$\pm$16 & 0.165$\pm$0.016 & NC1     \\
     & 3.52$\pm$0.35  & 0.62$\pm$0.06 &  -156$\pm$15 & 0.136$\pm$0.013 & C1  \\
     & 2.46$\pm$0.24  & 0.78$\pm$0.07 &  -154$\pm$18 & 0.078$\pm$0.007 & C2   \\
     & 1.02$\pm$0.10  & 0.87$\pm$0.08 &  -150$\pm$15 & 0.134$\pm$0.013 & C3   \\
     & 0.55$\pm$0.08  & 1.08$\pm$0.10 &  -132$\pm$11 & 0.653$\pm$0.065 & X       \\
1678 & 6.77$\pm$0.67  &    0$\pm$0    &     0$\pm$0  & 0.077$\pm$0.007 & core    \\
(Jan.~2014)& 2.52$\pm$0.25  & 0.24$\pm$0.02 &  -165$\pm$16 & 0.172$\pm$0.017 & NC1     \\
     & 1.60$\pm$0.16  & 0.53$\pm$0.05 &  -158$\pm$15 & 0.083$\pm$0.008 & X       \\
     & 3.48$\pm$0.34  & 0.72$\pm$0.07 &  -155$\pm$15 & 0.094$\pm$0.009 & C1  \\
     & 2.25$\pm$0.22  & 0.87$\pm$0.08 &  -152$\pm$15 & 0.165$\pm$0.016 & C2   \\
     & 0.87$\pm$0.08  & 1.03$\pm$0.10 &  -137$\pm$13 & 0.672$\pm$0.067 & C3   \\
1714 & 6.98$\pm$0.69  &    0$\pm$0    &     0$\pm$0  & 0.082$\pm$0.008 & core    \\
(Feb.~2014)& 2.59$\pm$0.25  & 0.36$\pm$0.03 &  -162$\pm$16 & 0.164$\pm$0.016 & NC1     \\
     & 1.65$\pm$0.16  & 0.60$\pm$0.06 &  -157$\pm$15 & 0.085$\pm$0.008 & X       \\
     & 3.36$\pm$0.33  & 0.76$\pm$0.07 &  -154$\pm$15 & 0.087$\pm$0.008 & C1  \\
     & 1.95$\pm$0.19  & 0.88$\pm$0.08 &  -151$\pm$15 & 0.164$\pm$0.016 & C2   \\
     & 0.61$\pm$0.06  & 1.13$\pm$0.11 &  -133$\pm$13 & 0.775$\pm$0.077 & C3   \\
1781 & 6.51$\pm$0.65  &    0$\pm$0    &     0$\pm$0  & 0.073$\pm$0.007 & core    \\
(May~2014)& 2.20$\pm$0.22  & 0.25$\pm$0.02 &  -172$\pm$17 & 0.123$\pm$0.012 & NC2     \\
     & 2.22$\pm$0.22  & 0.49$\pm$0.04 &  -161$\pm$16 & 0.096$\pm$0.009 & NC1     \\
     & 2.67$\pm$0.26  & 0.79$\pm$0.07 &  -154$\pm$15 & 0.115$\pm$0.011 & C1  \\
     & 3.20$\pm$0.32  & 0.94$\pm$0.09 &  -151$\pm$15 & 0.243$\pm$0.024 & C2   \\
     & 0.24$\pm$0.02  & 1.73$\pm$0.17 &  -148$\pm$14 & 0.242$\pm$0.024 & X       \\
1829 & 5.12$\pm$0.51  &    0$\pm$0    &     0$\pm$0  & 0.068$\pm$0.006 & core    \\
(Jun.~2014)& 2.47$\pm$0.24  & 0.28$\pm$0.02 &  -161$\pm$16 & 0.078$\pm$0.007 & NC2     \\
     & 2.21$\pm$0.22  & 0.52$\pm$0.05 &  -162$\pm$16 & 0.065$\pm$0.006 & NC1     \\
     & 2.03$\pm$0.20  & 0.77$\pm$0.07 &  -153$\pm$15 & 0.144$\pm$0.014 & C1  \\
     & 1.51$\pm$0.15  & 1.01$\pm$0.10 &  -151$\pm$15 & 0.176$\pm$0.017 & C2   \\
     & 0.60$\pm$0.06  & 1.27$\pm$0.12 &  -135$\pm$13 & 0.716$\pm$0.071 & C3   \\
1867 & 5.71$\pm$0.57  &    0$\pm$0    &     0$\pm$0  & 0.076$\pm$0.007 & core    \\
(Jul.~2014)& 2.09$\pm$0.20  & 0.29$\pm$0.02 &  -165$\pm$16 & 0.233$\pm$0.023 & NC2     \\
     & 2.11$\pm$0.21  & 0.50$\pm$0.05 &  -162$\pm$16 & 0.046$\pm$0.004 & X       \\
     & 1.86$\pm$0.23  & 0.68$\pm$0.06 &  -156$\pm$15 & 0.108$\pm$0.010 & NC1     \\
     & 1.51$\pm$0.15  & 0.83$\pm$0.08 &  -149$\pm$14 & 0.114$\pm$0.011 & C1  \\
     & 1.25$\pm$0.15  & 1.13$\pm$0.11 &  -152$\pm$15 & 0.257$\pm$0.025 & C2   \\
     & 0.42$\pm$0.04  & 1.43$\pm$0.14 &  -129$\pm$12 & 0.733$\pm$0.073 & C3   \\
1924 & 5.80$\pm$0.58  &    0$\pm$0    &     0$\pm$0  & 0.056$\pm$0.005 & core    \\
(Sep.~2014)& 1.89$\pm$0.18  & 0.25$\pm$0.02 &  -157$\pm$15 & 0.137$\pm$0.013 & X       \\
     & 1.78$\pm$0.17  & 0.52$\pm$0.05 &  -159$\pm$19 & 0.048$\pm$0.004 & NC2     \\
     & 2.44$\pm$0.24  & 0.74$\pm$0.09 &  -156$\pm$15 & 0.098$\pm$0.009 & NC1     \\
     & 1.65$\pm$0.16  & 0.87$\pm$0.08 &  -148$\pm$18 & 0.176$\pm$0.017 & C1  \\
     & 0.73$\pm$0.07  & 1.20$\pm$0.12 &  -151$\pm$15 & 0.198$\pm$0.019 & C2   \\
     & 0.47$\pm$0.04  & 1.52$\pm$0.15 &  -130$\pm$20 & 0.713$\pm$0.071 & C3   \\
1976 & 4.59$\pm$0.45  &    0$\pm$0    &     0$\pm$0  & 0.059$\pm$0.005 & core    \\
(Nov.~2014)& 0.83$\pm$0.08  & 0.23$\pm$0.02 &  -159$\pm$15 & 0.107$\pm$0.010 & X       \\
     & 1.10$\pm$0.11  & 0.53$\pm$0.05 &  -159$\pm$19 & 0.066$\pm$0.006 & NC2     \\
     & 2.45$\pm$0.56  & 0.79$\pm$0.07 &  -154$\pm$15 & 0.147$\pm$0.014 & NC1     \\
     & 0.69$\pm$0.06  & 1.06$\pm$0.10 &  -149$\pm$14 & 0.258$\pm$0.025 & C1  \\
     & 0.51$\pm$0.05  & 1.54$\pm$0.15 &  -134$\pm$17 & 0.935$\pm$0.093 & C3   \\
1997 & 5.90$\pm$0.59  &    0$\pm$0    &     0$\pm$0  & 0.068$\pm$0.006 & core    \\
(Dec.~2014)& 1.33$\pm$0.13  & 0.16$\pm$0.01 &  -162$\pm$16 & 0.115$\pm$0.011 & NC3     \\
     & 0.72$\pm$0.07  & 0.48$\pm$0.04 &  -160$\pm$16 & 0.050$\pm$0.009 & X       \\
     & 0.95$\pm$0.09  & 0.68$\pm$0.06 &  -158$\pm$15 & 0.108$\pm$0.010 & NC2     \\
     & 2.31$\pm$0.23  & 0.84$\pm$0.08 &  -154$\pm$15 & 0.148$\pm$0.014 & NC1     \\
     & 0.53$\pm$0.05  & 0.97$\pm$0.09 &  -147$\pm$19 & 0.198$\pm$0.019 & C1  \\
     & 0.27$\pm$0.05  & 1.28$\pm$0.11 &  -150$\pm$15 & 0.236$\pm$0.023 & C2   \\
     & 0.32$\pm$0.03  & 1.61$\pm$0.14 &  -127$\pm$19 & 0.667$\pm$0.066 & C3   \\
     & 0.08$\pm$0.02  & 1.65$\pm$0.19 &  -153$\pm$15 & 0.325$\pm$0.032 & X       \\\hline
\end{tabular}\\
JD$^{\prime}$ = JD - 2454000, \\
S : The integrated flux density in the component, \\
r : The radial distance of the component center from the center of the map, \\
$\theta$ : The position angle of the center of the component, and \\ 
$\phi$ : The FWHM of the component. \\
Comp$^{a}$ :  Identification of the individual components. If a component appeared only in a single epoch, 
we labeled it with X.  \\
\end{center}
\label{tab1}
\end{table}

\end{appendix}

\end{document}